\documentclass[a4paper,twoside]{article}
\newif\ifdraft \global\drafttrue

\usepackage[T1]{fontenc}
\usepackage{a4wide}
\usepackage{times}
\usepackage{graphicx}
\usepackage{theorem}
\usepackage[colorlinks=true,hyperindex=true]{hyperref}
\usepackage{color}
\usepackage{fancyhdr}
\usepackage{amsmath}
\usepackage{bbm}
\usepackage{amsfonts}
\usepackage{enumerate}
\usepackage{cancel}
\ifdraft
\fi
\newtheorem{theorem}{Theorem}[section]
\newtheorem{proposition}[theorem]{Proposition}
\newtheorem{lemma}[theorem]{Lemma}
\newtheorem{definition}[theorem]{Definition}
\newtheorem{corollary}[theorem]{Corollary}
\theoremstyle{plain}
\theorembodyfont{\upshape}

\newcounter{smallarabics}
\newenvironment{arabicenumerate}
{\begin{list}{{\normalfont\textrm{(\arabic{smallarabics})}}}
{\usecounter{smallarabics}\setlength{\itemindent}{0cm}
\setlength{\leftmargin}{5ex}\setlength{\labelwidth}{4ex}
\setlength{\topsep}{0.75\parsep}\setlength{\partopsep}{0ex}
\setlength{\itemsep}{0ex}}}
{\end{list}}

\def\bel{\begin{lemma}}
\def\eel{\end{lemma}}
\def\bec{\begin{corollary}}
\def\eec{\end{corollary}}
\def\bet{\begin{theorem}}
\def\eet{\end{theorem}}
\def\bed{\begin{definition}}
\def\eed{\end{definition}}
\def\bep{\begin{proposition}}
\def\eep{\end{proposition}}
\def\ben{\begin{arabicenumerate}}  
\def\een{\end{arabicenumerate}}
\def\beq{\begin{equation}}
\def\eeq{\end{equation}}  
\def\demo{{\noindent\bf Proof.\ }}
\def\qed{\hfill$\square$}
\newcommand{\e}{\mathrm{e}}
\let\oldi\i
\newcommand{\myi}{\oldi}
\renewcommand{\i}{\mathrm{i}}

\renewcommand{\d}{\mathrm{d}}

\def\rr{{\mathbb R}}
\def\cc{{\mathbb C}}
\def\one{{\mathbbm 1}}
\def\cJ{\mathcal{J}}
\def\cI{\mathcal{I}}
\def\cL{\mathcal{L}}
\def\cM{\mathcal{M}}

\def\cB{\mathcal{B}}
\def\cS{\mathcal{S}}
\def\cT{\mathcal{T}}
\def\cP{\mathcal{P}}

\def\cK{\mathcal{K}}
\def\cH{\mathcal{H}}

\def\cR{\mathcal{R}}
\def\cO{\mathcal{O}}
\def\bPt{\bar P^{\kern1pt t}}
\def\bQt{\bar Q^{\kern1pt t}}

\def\CP{{\rm CP}}

\def\fS{\mathfrak{S}}

\def\Re{\mathrm{Re}\,}

\def\Ker{\mathrm{Ker}\,}

\def\sp{\mathrm{sp}}

\def\eq{\mathrm{eq}}

\def\tr{\mathrm{tr}}

\def\Ent{{\rm Ent}}

\def\cB{{\cal B}}
\def\semigroup{{\{\e^{t\cL}\}_{t\geq 0}}}

\def\bar{\overline}

\def\cal{\mathcal}
\def\one{{\mathbbm 1}}
\def\KMSB{{\rm KMS}${}_\beta$}

\begin{document}
\thispagestyle{empty}
\title{\bf\Huge Entropic fluctuations of\\ quantum dynamical semigroups\\ \vskip1cm}
\author{\sc V. Jak\v{s}i\'c$^{a}$, C.-A. Pillet$^{a,b}$, M. Westrich$^{a}$\\ \\ \\
\\ \\ \\ 
$^a$Department of Mathematics and Statistics\\ 
McGill University\\
805 Sherbrooke Street West \\
Montreal,  QC,  H3A 2K6, Canada
 \\ \\
$^b$
Aix-Marseille Universit\'e, CNRS, CPT, UMR 7332, Case 907, 13288 Marseille, France\\
Universit\'e de Toulon, CNRS, CPT, UMR 7332, 83957 La Garde, France\\
FRUMAM
}
\def\today{}
\maketitle

% Keywords: nonequilibrium statistical mechanics, entropy production, 
% Evans-Searless fluctuation relation, Lindbladian, linear response theory,
% weak coupling limit, Markovian quantum dynamics, full counting statistics
% detailed balance, 

% HAL: paperId: hal-00823948  password: re106t
% mp_arc:TyQTk

\vskip 1cm
\begin{quote}
{\bf Abstract.} 
We study a class of finite dimensional quantum dynamical semigroups $\{\e^{t\cL}\}_{t\geq 0}$
whose generators
$\cL$ are sums of Lindbladians satisfying the detailed balance condition. Such semigroups
arise in the weak coupling (van Hove) limit of Hamiltonian dynamical systems describing
open quantum systems out of equilibrium. We prove a general entropic fluctuation theorem
for this class of semigroups by relating the cumulant  generating function of entropy transport to the
spectrum of a family of deformations of the generator ${\cL}$. We show that, besides the celebrated
Evans-Searles symmetry, this cumulant generating function also satisfies the translation symmetry
recently discovered by Andrieux {\sl et al.}, and that in the linear regime near equilibrium
these two symmetries yield Kubo's and Onsager's linear response relations.

\end{quote}
\vskip 1cm
\begin{center}
{\Large\bf Dedicated to Herbert Spohn on the occasion of his 65th birthday.}
\end{center}
%%%%%%%%%%%%%%%%%%%%%%%%%%%%%%%%%%%%%%%%%%%%%%%
%%%%%%%%%%%%%%%%%%%%%%%%%%%%%%%%%%%%%%%%%%%%%%%
\section{Introduction}
%%%%%%%%%%%%%%%%%%%%%%%%%%%%%%%%%%%%%%%%%%%%%%%
%%%%%%%%%%%%%%%%%%%%%%%%%%%%%%%%%%%%%%%%%%%%%%%

Markov semigroups are widely used to model non-equilibrium phenomena in classical statistical
physics. Their non-commutative counterparts ----  quantum dynamical semigroups --- play the
same role in quantum statistical physics (see, e.g., \cite{AL,Re} for pedagogical introductions to the
subject). The development of the mathematical theory of QDS started in 1974 with the seminal works of Brian
Davies \cite{Da1, Da2, Da3} where he showed that QDS emerge as effective dynamics of open systems weakly
coupled to extended reservoirs. These groundbreaking works were followed by the celebrated 1976 papers
of Lindblad \cite{Li1, Li2} and Gorini, Kossakowski and Sudarshan \cite{GKS} on the structure of the generator of
QDS (see also \cite{CE}). Herbert made several fundamental contributions at this early stage of development.
In \cite{Sp1, Sp2} he gave efficient criteria for the existence and uniqueness of a stationary state and approach to
equilibrium. Together with Joel Lebowitz, in \cite{LS1} he developed a comprehensive picture of the nonequilibrium
thermodynamics of weakly coupled open systems. This work remains a standard reference and has
been a source of inspiration for many later developments on the subject, including the present one. Among
other things, in \cite{LS1} Herbert and Joel introduced the central concept of entropy production, which was
further discussed in \cite{Sp3}, and developed the linear response theory for thermodynamical forces. The
closely related linear response theory for weakly coupled open systems under mechanical drive was developed
by Herbert in a joint paper with Brian Davies \cite{DS}. In another enlightening work, Herbert and R.~D\"umcke \cite{D\"uSp}
showed that some of the generators that were (and sometimes still are) used to describe
the weak coupling limit lead to negative probabilities. Years later, Herbert came back to the subject and,
with Walter Aschbacher, showed that when properly applied to nonequilibrium situations, the algebraic
criterion of \cite{Sp2} also ensures the strict positivity of entropy production \cite{AS}.

In 1993/4, using a new scheme to construct nonequilibrium statistical ensembles of interacting particle
systems, Evans, Cohen and Morriss discovered some universal features of the fluctuations of entropy production
in transient regimes of deterministic classical systems out of thermal equilibrium \cite{ECM, ES}. One
year later, Gallavotti and Cohen proved that some steady states of highly chaotic dynamical systems (SRB
measures of transitive Anosov systems) display the same features \cite{GC1, GC2}. These discoveries, nowadays
called fluctuation relations or fluctuation theorems, triggered a large amount of works during the last
two decades (see \cite{RM, JPR} and references therein). In particular, Kurchan showed that the fluctuation
relations hold for a Brownian particle in a force field \cite{Ku1}. Subsequently, Herbert and Joel formulated
and proved fluctuation relations for general Markov processes \cite{LS2} while Maes derived a local version of
the fluctuation relations from the Gibbsian nature of the path space measure associated to such processes
\cite{M} (see also \cite{MRV}). As shown by Gallavotti \cite{Ga}, the fluctuation relations can be seen as a far from
equilibrium generalization of the familiar near equilibrium fluctuation-dissipation relations (Green-Kubo
formulae, Onsager reciprocity relations).

The attempts to extend fluctuation relations to quantum domain have led to a number of surprises. The naive
quantization of the classical transient fluctuation relations fails and there is no obvious way to implement
the steady state fluctuation relations. These problems have attracted a lot of interest and generated a huge
literature which we will not try to review here. We shall only mention a few works which, in our opinion,
are relevant to the development of a mathematical understanding of the subject. The interested reader can
consult \cite{EHM} for an exhaustive review and an extended list of references to the physics literature and
\cite{JOPP, JP} for recent mathematical developments.

To our knowledge, a (transient) quantum fluctuation relation based on operationally defined counting statistics
was first derived by Kurchan in 2000 \cite{Ku2}. Shortly afterwards, Matsui and Tasaki obtained an apparently unrelated abstract
fluctuation relation for open quantum systems in terms of the spectral measure of a relative modular
operator \cite{MT}. The connection between their result and the counting statistics of entropic transport was
established in \cite{JOPP}.

Within the framework of QDS, de Roeck and Maes \cite{dRM} used the unraveling technique to obtain the
first complete transient fluctuation theorem (see Section 6). The relation between this Markovian approach
to fluctuations and the Hamiltonian description of the dynamics of a small system weakly coupled to an
extended environment was discussed by de Roeck in \cite{dR1} and by Derezi\'nski, de Roeck and Maes 
\cite{DdRM} (see also Section 5).  The works \cite{dRM, dR1, DdRM} complete the program of \cite{LS1} regarding 
nonequilibrium thermodynamics of weakly coupled open systems. The first proof of the transient fluctuation theorem for a fully Hamiltonian system
(the spin-boson model) was given by de Roeck in the important  paper \cite{dR2} (see also \cite{dRK1, dRK2,
JPPW}). 

Among the non-rigorous works let us mention the important observation of Andrieux, Gaspard, Monnai 
and Tasaki \cite{AGMT}  that global conservation laws (energy and charge) induce translation
symmetries in the cumulant generating function of (energy and charge) fluxes. Translation symmetries
and entropic fluctuation relation lead to fluctuation relations for individual fluxes and, following the 
arguments of \cite{Ga, LS2}, to Green-Kubo and Onsager relations near thermal equilibrium.
We also mention the recent work by Chetrite and Mallik \cite{CM} where finite time fluctuation relations
(similar to the classical transient Evans-Searles relations) for time-reversible quantum dynamical 
semigroups are derived.

This work is of a review nature and we do not prove any specific new results. The purpose of the paper is to provide an abstract general setup for the non-equilibrium statistical mechanics
of QDS and to generalize and streamline the proof of the full fluctuation theorem of \cite{dRM, dR1, DdRM} emphasizing (in the
spirit of \cite{Sp3}) the minimal mathematical structure behind the result. 
The fluctuation theorem we discuss 
includes large deviation bounds and the central limit theorem for individual entropic fluxes, as well as 
linear response formulae and the fluctuation-dissipation relations near equilibrium, and applies  to the
weakly coupled quantum systems studied in \cite{LS1}. Although the paper is 
mathematically  self-contained, it is intended for readers familiar with the works \cite{Sp1, Sp2, Sp3, LS1}. This paper can be also 
viewed as an introduction to \cite{JPPW}  where we discuss fluctuation relations  and 
non-equilibrium statistical mechanics of the fully Hamiltonian Pauli-Fierz systems.

The paper is organized as follows. In Section \ref{prelim} we recall basic definitions and facts about positive maps
and QDS. In Section \ref{sect-qds} we introduce the setup of QDS out of equilibrium, and state our main results. In
Section \ref{sect-WCL} we show that open systems weakly coupled to thermal reservoirs fit into our general setup. In
Sections \ref{sect-FCS}  and \ref{sec-unravel} we relate our results to the full counting statistics of entropic transport and the unraveling of
quantum dynamical semigroups. Finally, Section \ref{sect-proofs}  is devoted to the proofs.

\bigskip

{\noindent \bf Acknowledgment.} The research of V.J. was partly supported by NSERC. The research of C.-A.P. was
partly supported by ANR (grant 09-BLAN-0098). C.-A.P. is also grateful to the Department of Mathematics
and Statistics at McGill University and to CRM (CNRS - UMI 3457) for hospitality and generous
support during his stay in Montreal where most of this work was done. We are grateful to  J.~Derezi\'nski,  
B.~Landon, and A.~Panati for useful comments. We also thank C.~Maes and W.~de Roeck for interesting
related discussions.

%%%%%%%%%%%%%%%%%%%%%%%%%%%%%%%%
\section{Preliminaries}
\label{prelim}

Let $\cH$ be a finite dimensional Hilbert space and $\cO=\cB(\cH)$ the $C^\ast$-algebra of all linear 
operators on $\cH$ (the identity operator will be always denoted by $\one$). Equipped with the inner 
product $\langle X|Y \rangle=\tr(X^\ast Y)$, $\cO$ is a Hilbert space. The adjoint and the spectrum of 
a linear map  $\Phi:\cO\rightarrow\cO$  are denoted by $\Phi^\ast$ and $\sp (\Phi)$. ${\rm Id}$ 
denotes the identity of $\cB(\cO)$. A subset ${\cal A}\subset {\cal O}$ is called self-adjoint if 
$X\in {\cal A} \Rightarrow X^\ast \in {\cal A}$. The commutant of a subset ${\cal A}\subset \cO$ is 
${\cal A}^\prime=\{B\in\cO\,|\,AB=BA\text{ for all }A\in\mathcal{A}\}$.

We denote by $\cO_+=\{X\,|\, X\geq 0\}$ the cone of positive elements of  $\cO$. A linear map $\Phi$ 
from  $\cO$ to another unital $C^\ast$-algebra $\cB$ is called unital if $\Phi(\one)=\one$, positive if  
$\Phi(\cO_+)\subset  \cB_+$, and positivity improving if $\Phi(X)>0$ for all non-zero $X\in  \cO_+$. 
A positive linear map is automatically a $\ast$-map, {\sl i.e.}, it satisfies
$\Phi(X^\ast)=\Phi(X)^\ast$. A positive linear map $\Phi: \cO \rightarrow \cO$ is called Schwartz if 
\[
\Phi(X^\ast)\Phi(X)\leq \|\Phi\|\Phi(X^\ast X),
\]
for all $X\in \cO$. Note that if $\Phi$ is Schwartz, then $\|\Phi\|=\|\Phi(\one)\|$. 

A state on ${\cal O}$ is a positive and unital linear map $\rho: \cO \rightarrow \cc$. Any state $\rho$ has the form 
$\rho(X)=\tr(DX)$ for some $D\in \cO_+$ satisfying $\tr(D)=1$. Such an operator $D$ is called a density matrix. In the following, 
we shall use the same symbol to denote a density matrix and the state it induces on $\cO$ (hence,
$\rho(X)=\tr(\rho X)$, etc.). With this convention, the set of states 
on $\cO$, which we denote by ${\mathfrak S}$, is a closed convex subset of $\cO_+$. A state $\rho$ is called faithful 
if $\rho>0$, and we denote by ${\mathfrak S}_{\rm f}$ the set of faithful states. ${\mathfrak S}_{\rm f}$ is an open convex and dense 
subset of ${\mathfrak S}$.

A linear map $\Phi: \cO \rightarrow \cO$ is called completely positive (CP) if $\sum_{i, j}B_i^\ast \Phi(A_i^\ast A_j)B_j\geq 0$ 
for any finite families $\{A_1, \cdots, A_N\}, \{B_1, \cdots, B_N\}\subset \cO$. Equivalently, $\Phi$ is CP if 
$\Phi \otimes {\rm Id}$ is a positive map on $\cO\otimes \cB(\cc^N)$ for all $N\geq 1$.   A CP map 
is automatically Schwartz. We denote by $\CP(\cO)$ the monoid of 
completely positive maps, and by $\CP_1(\cO)$ the sub-monoid of unital maps. $\CP(\cO)$ is a convex cone and 
$\CP_1(\cO)$ is a convex set. Stinespring's theorem \cite{St} asserts that $\Phi \in \CP(\cO)$ iff there exists a finite 
family $\{V_j\}_{j\in J}$ in $\cO$ such that 
\begin{equation}
\label{prelim-1}
\Phi(X)=\sum_{j\in J} V_j^\ast XV_j,
\end{equation}
for all $X\in \cO$. The formula (\ref{prelim-1}) is called a Kraus representation of $\Phi$. Such representation is in general not unique.

Unital CP maps naturally arise in the quantum mechanics of open systems. Indeed, assume that the 
quantum system $\cS$ with Hilbert space $\cH$ interacts with some environment described by 
the Hilbert space $\cH_{\rm env}$. According to the general structure of quantum mechanics, 
the evolution of the joint system over some time interval is given by a unitary $U$ on 
$\cH\otimes\cH_{\rm env}$. Thus, if $X$ is an observable of the system $\cS$, then its Heisenberg
evolution over the considered time interval is given by the map
$$
\Phi(X)=\tr_{\cH_{\rm env}}\left((\one\otimes\rho_{\rm env})U^\ast (X\otimes \one)U\right),
$$
where $\tr_{\cH_{\rm env}}(\,\cdot\,)$ denotes the partial trace over the environment Hilbert space
and $\rho_{\rm env}$ is the  initial state of the environment. One easily checks that $\Phi$ is a unital 
CP map such that, for any state $\rho$ of $\cS$,
$$
\tr(\rho\Phi(X))=\tr\left((\rho\otimes\rho_{\rm env})U^\ast (X\otimes \one)U\right).
$$

A positive linear map $\Phi$ is called irreducible (in the sense of Davies \cite{Da0}) if the inequality 
$\Phi({P})\leq \lambda P$, where $P$ is a projection and $\lambda>0$, holds only for   $P=0$ or $P=\one$. 
If $\Phi$ is positivity improving, then obviously $\Phi$ is irreducible. In terms of a Kraus decomposition, 
 irreducibility can be characterized as follows (see, {\sl e.g.}, \cite{Sc}):
\bet\label{prelim-thm1}
Let $\Phi$ be a $\CP(\cO)$ map with a Kraus decomposition (\ref{prelim-1}) and  let ${\cal A}$ be the subalgebra of 
${\cal O}$ generated  by $\{V_j\,|\, j\in J\}$ and $\one$. Then $\Phi$ is irreducible iff ${\cal A}\psi=\cH$ for any 
non-zero vector $\psi \in \cH$.
\eet
 
For reader's convenience, we shall prove Theorem~\ref{prelim-thm1} in Section \ref{sect-AQDS}.

The adjoint $\Phi^\ast$ of a linear map $\Phi$ is positive/positivity improving/CP/irreducible iff $\Phi$ is. 
$\Phi^\ast$ is trace preserving, {\sl i.e.}, $\tr (\Phi^\ast(X))=\tr(X)$ for all $X\in \cO$, iff $\Phi$ is unital. In particular, 
$\Phi^\ast$ maps ${\mathfrak S}$ into itself iff $\Phi$ is positive and unital. A state  $\rho \in {\mathfrak S}$ is called 
$\Phi$-invariant if $\Phi^\ast(\rho)=\rho$, which is equivalent to $\rho(\Phi(X))=\rho(X)$ for all $X\in \cO$.

Let $\semigroup$ be a continuous semigroup of linear maps on $\cO$ generated by a linear map $\cL$. This semigroup is 
called unital/positive/positivity improving/$\CP(\cO)$/$\CP_1(\cO)$ iff $\e^{t \cL}$ is for all $t>0$. A $\CP_1(\cO)$ semigroup is 
called {\em quantum dynamical semigroup} (QDS).\footnote{The name {\em quantum Markov semigroup} is also used in 
the literature.} 

Let $\semigroup$ be a positive unital semigroup. A state $\rho$ is called steady (or stationary) if 
$\rho(\e^{t\cL}(X))=\rho(X)$ for all $t\geq 0$ and $X\in \cO$. Clearly, $\rho$ is steady iff $\cL^\ast(\rho)=0$. 

A positive unital semigroup $\semigroup$ is said to be relaxing to a steady state $\rho_+$ if 
\begin{equation}\label{prelim-2}
\lim_{t\rightarrow \infty}\e^{t \cL^\ast}(\rho)=\rho_+,
\end{equation}
for all $\rho \in {\mathfrak S}$. The relaxation is exponentially fast if there exists $\gamma>0$ such that for all 
states $\rho$, 
\[
\e^{t \cL^\ast}(\rho)=\rho_++ O(\e^{-\gamma t}),
\]
as $t\rightarrow \infty$. The relaxation to a steady state is an ergodic property that plays a fundamental role 
in the statistical mechanics of QDS. 

Our study of the large deviation theory of QDS will be based on the following result. 
\bet\label{prelim-thm2}
Let $\semigroup$ be a positivity improving $\CP(\cO)$ semigroup and 
\[
\ell=\max\{ \Re \lambda\,|\, \lambda \in \sp(\cL)\}.
\]
Then $\ell$ is a simple eigenvalue of $\cL$ and is the only eigenvalue of $\cL$ on the line $\Re z=\ell$. 
For any state $\rho$ on $\cO$, one has 
\begin{equation}\label{prelim-3}
\ell=\lim_{t\rightarrow \infty}\frac{1}{t}\log \rho(\e^{t \cL}(X)),
\end{equation}
for all non-zero $X\in \cO_+$. If in addition the semigroup $\semigroup$ is unital, then $\ell=0$ and the 
semigroup is relaxing exponentially fast to a faithful steady state $\rho_+$.
\eet
The proof of this theorem is based on the Perron-Frobenius theory of positive maps developed 
in \cite{EHK} and is given in Section \ref{sect-AQDS}.

It is a fundamental result of Lindblad \cite{Li1, Li2}, Gorini, Kossakowski and Sudarshan \cite{GKS}, 
and Christensen and Evans \cite{CE}, that $\semigroup$ is a $\CP(\cO)$ semigroup iff there are 
$K\in \cO$ and $\Phi \in \CP(\cO)$ such that 
\begin{equation}\label{prelim-4}
\cL(X)= K^\ast X + XK +\Phi(X),
\end{equation}
for all $X\in \cO$. For short, we shall call the generator of a $\CP(\cO)$ semigroup a {\em Lindbladian}, and 
the r.h.s.\;of Eq. (\ref{prelim-4}) a {\em Lindblad decomposition} of $\cL$. Although the Lindblad decomposition is not 
unique, it can be effectively used to characterize some important properties of the semigroup. In particular, 
we have:
\bet\label{prelim-thm3} Let $\semigroup$ be a $\CP(\cO)$ semigroup and $\cL(X)= K^\ast X + XK +\Phi(X)$ a 
Lindblad decomposition. If $\Phi$ is irreducible, then the semigroup is positivity improving. 
\eet
We shall prove this theorem in Section \ref{sect-AQDS}. Theorems~\ref{prelim-thm1} and~\ref{prelim-thm3} provide an effective 
criterion for verifying the positivity improving assumption of Theorem~\ref{prelim-thm2} (see Section \ref{sect-WCL}). 

If $\semigroup$ is a QDS, then $\cL(\one)=0$, and it follows from (\ref{prelim-4}) that 
\begin{equation}\label{prelim-5}
\cL(X)=\i [T, X]-\frac{1}{2}\{ \Phi(\one), X\} + \Phi(X),
\end{equation}
where $T$ is a self-adjoint element of $\cO$ and $\Phi \in \CP(\cO)$. We shall also refer to the r.h.s.\;of 
Eq. (\ref{prelim-5}) as a {\em Lindblad decomposition} of $\cL$.

The {\em dissipation function} of a QDS $\semigroup$ is the sesquilinear map $D: \cO\times \cO \rightarrow \cO$ defined by 
\[D(X,Y)=\cL(X^\ast Y)-\cL(X^\ast)Y-X^\ast\cL(Y).
\]
If (\ref{prelim-5}) is the Lindblad decomposition of $\cL$ and 
(\ref{prelim-1}) a Kraus decomposition of $\Phi$, then 
\[D(X, X)=\sum_{j\in J}[V_j, X]^\ast[V_j, X].
\]
Hence, $D(X, X)\geq 0$ and $D(X, X)=0$ iff  $X\in\{V_j\,|\, j\in J\}^\prime$.  The dissipation function of a QDS was  introduced 
by Lindblad in \cite{Li1} and has played an important role in many subsequent works on the subject.  

The {\em detailed balance condition} and  {\em time-reversal invariance} will play an important role in our work. 
Both properties refer to a pair $(\rho, \cL)$, where $\rho$ is a faithful state and $\cL$ is the generator of a  QDS. Note that 
any faithful state  induces an inner product $\langle X|Y\rangle_\rho=\langle X\rho^{1/2}|Y\rho^{1/2}\rangle=
\tr (\rho X^\ast Y)$ on $\cO$. We call the  $\rho$-adjoint of a linear map $\Phi$ its adjoint $\Phi^\rho$ w.r.t.\;this inner product. In particular, 
we say that a linear map $\Phi$ is $\rho$-self-adjoint if $\Phi^\rho=\Phi$. 

\begin{definition}\label{prelim-def}
Consider a pair $(\rho, \cL)$, where $\rho$ is a faithful state and $\cL$ is a Lindbladian generating a QDS.
\begin{enumerate}[{\rm (a)}]
\item The pair $(\rho, \cL)$ is said satisfy the detailed balance condition if $\cL^\ast(\rho)=0$ and there exists Lindblad 
decomposition $\cL=\i [T, \,\cdot\,]-\frac{1}{2}\{\Phi(\one),\,\cdot\,\}+\Phi$ such that $\Phi$ is $\rho$-self-adjoint.

\item The pair $(\rho, \cL)$ is said to be  time-reversal invariant (TRI) if there exists an involutive anti-linear $\ast$-automorphism 
$\Theta: \cO \rightarrow \cO$, called the time-reversal,   such that $\cL^\rho\circ \Theta=\Theta \circ \cL$ and $\Theta(\rho)=\rho$.  
\end{enumerate}
\end{definition}

Definition \ref{prelim-def} (a) is equivalent to the definition of detailed balance given by Kossakowski, Frigerio, 
Gorini, and Verri \cite{KFGV} (see Theorem~\ref{utah-knee} below). The above definition, however, is technically and conceptually 
more suitable for our purposes.\footnote{Alternative  definitions of detailed balance 
can be found in \cite{Ag, Al}.} The detailed balance condition is characteristic of QDS describing the interaction of a system 
$\cS$ with an environment at equilibrium (see \cite{KFGV, LS1}).

For the motivation regarding the definition of time-reversal we refer the reader to Section \ref{sect-WCL} and 
\cite{Ma, FU}.  We recall that $\Theta: \cO \rightarrow \cO$ is an involutive anti-linear $\ast$-automorphism iff there exists an 
anti-unitary involution $\theta: \cH\rightarrow \cH$ such that $\Theta(X)=\theta X\theta$ (see Exercise 4.36 in \cite{JOPP}), and 
that $\Theta(\rho)=\rho$ iff $\rho(\Theta(X))=\rho(X^\ast)$ for all  $X\in\cO$.

%%%%%%%%%%%%%%%%%%%%%%%%%%%%%%%%%%%%%%%%%%%%%%%
%%%%%%%%%%%%%%%%%%%%%%%%%%%%%%%%%%%%%%%%%%%%%%%
\section{Quantum dynamical semigroups out of equilibrium}
\label{sect-qds}
%%%%%%%%%%%%%%%%%%%%%%%%%%%%%%%%%%%%%%%%%%%%%%%
%%%%%%%%%%%%%%%%%%%%%%%%%%%%%%%%%%%%%%%%%%%%%%%
\subsection{The setup}
\label{set-setup}
We shall study QDS $\semigroup$ on $\cO=\cB(\cH)$, $\dim \cH <\infty$, satisfying the following
ergodicity condition 

\begin{quote}
{\bf (ER)} The QDS $\semigroup$ is positivity improving. 
\end{quote}

Furthermore, we shall focus on semigroups whose generator $\cL$ has the special structure given by 

\begin{quote}
{\bf (DB)} There are $M$ pairs $(\rho_j, \cL_j)$ satisfying the detailed balance condition such that 
\[\cL=\sum_{j=1}^M \cL_j.\] 
\end{quote}

We shall interpret each of the $M$ {\sl sub-Lindbladians} $\cL_j$ as describing the
interaction of a quantum system $\cS$ with some reservoir $\cR_j$.
A QDS is out of equilibrium if Hypothesis (DB) holds and $\rho_j\not=\rho_k$ for some
pair $j,k$ (such a QDS describes the interaction of the system $\cS$ with an environment
$\sum_j \cR_j$ out of equilibrium).

A QDS satisfying (DB) is called time-reversible if  

\begin{quote}
{\bf (TR)}  All the pairs $(\rho_j, \cL_j)$ are time-reversal invariant with the same time-reversal map $\Theta$. 
\end{quote}

Finally, we shall investigate more closely the special case where the states $\rho_j$ are thermal
equilibrium states at inverse temperature $\beta_j>0$ for the same Hamiltonian.

\begin{quote}
({\bf KMS}${}_{\boldsymbol{\beta}}$) $\beta=(\beta_1,\ldots,\beta_M)\in\rr_+^M$ and there exists a 
self-adjoint element $H_\cS\in\cO$ such that 
$$
\rho_j=\frac{\e^{-\beta_jH_\cS}}{\tr(\e^{-\beta_jH_\cS})},
$$
for $j=1,\ldots,M$.
\end{quote}

As we shall see in Section~\ref{sect-WCL}, Hypotheses  (ER), (DB) and (\KMSB) are  naturally satisfied by the QDS
describing the weak coupling (van Hove) limit  dynamics of an open quantum system $\cS$ with Hilbert space 
$\cH$  interacting with an environment
made of $M$ thermal reservoirs. In this case, the Lindbladian $\cL_j$ pertains to the interaction
of $\cS$ with the $j^{\rm th}$ reservoir and the state $\rho_j$ is a steady state
of the system coupled only to this reservoir. If the joint dynamics of the system and reservoirs
is time-reversal invariant, then Hypothesis  (TR) is also  satisfied. 

\subsection{Main result}
Suppose that a QDS $\semigroup$ satisfies Hypothesis (DB). 
Our main technical result concerns the properties of the deformations
of this  QDS  generated by
\beq
\cL_{(\alpha)}(X)=\sum_{j=1}^M\cL_j(X\rho_j^{-\alpha_j})\rho_j^{\alpha_j},
\label{cLalphadef}
\eeq
where $\alpha=(\alpha_1,\ldots,\alpha_M)\in\rr^M$. We will use the notation
$1=(1,\ldots,1)$ wherever the  meaning is clear within the context, {\sl e.g.,}
$1-\alpha=(1-\alpha_1,\ldots,1-\alpha_M)$. Let 
\[e(\alpha)=\max\{ \Re \lambda\,|\, \lambda \in \sp(\cL_{(\alpha)})\}.
\]

\bet\label{thm-1} Suppose that Hypothesis (DB) holds. Then: 
\begin{enumerate}[{\rm (1)}]
\item $\{\e^{t\cL_{(\alpha)}}\}_{t\geq 0}$ is a $\CP(\cO)$ semigroup for all $\alpha\in\rr^M$.
\item For any state $\rho$ on $\cO$, there is a Borel probability measure $P_\rho^t$ on $\rr^M$ such that
$$
\tr \left(\rho\e^{t\cL_{(\alpha)}}(\one)\right)=\int_{\rr^M}\e^{-t\alpha\cdot\varsigma}\d P_\rho^t(\varsigma).
$$
We denote by $\langle\,\cdot\,\rangle_{\rho,t}$ the expectation w.r.t.\;this measure.
\end{enumerate}
In the remaining statements we assume that Hypothesis (ER) is satisfied. 
\begin{enumerate}[{\rm (1)}]
\setcounter{enumi}{2}
\item For all $\alpha \in \rr^M$ the $\CP(\cO)$ semigroup $\{\e^{t\cL_{(\alpha)}}\}_{t\geq 0}$ is positivity improving. 
In particular, the QDS $\semigroup$ is relaxing exponentially fast to a steady state $\rho_+$. 

\item 
For all $\alpha \in \rr^M$, $e(\alpha)$ is a simple eigenvalue of $\cL_{(\alpha)}$ and this operator has 
no other eigenvalues on the line $\Re z = e(\alpha)$. Moreover, for any state $\rho$ and all 
$\alpha \in \rr^M$, 
\beq
\lim_{t\to\infty}\frac1t\log\left<\e^{-t\alpha\cdot\varsigma}\right>_{\rho,t}=e(\alpha).
\label{eq-elim}
\eeq
\item The function $\rr^M\ni\alpha\mapsto e(\alpha)$ is real analytic and convex.
\item Relation~\eqref{eq-elim} holds for $\alpha$ in an open neighborhood of $\rr^M$ in $\cc^M$.
\item If Hypothesis  {\rm (TR)} is satisfied, then 
\beq\label{es-snow}
e(1-\alpha)=e(\alpha),
\eeq
 for all $\alpha\in\rr^M$.
\item If Hypothesis {\rm (\KMSB)} is satisfied, then 
\[
e(\alpha+\lambda\beta^{-1})=e(\alpha),
\]
for all $\alpha\in\rr^M$ and all $\lambda\in\rr$ with $\beta^{-1}=(\beta_1^{-1},\ldots,\beta_M^{-1})$.
\end{enumerate}
\eet
{\bf Remark 1.} The identity  (\ref{es-snow}) is the QDS analog of the generalized Evans-Searles symmetry of time-reversal 
invariant classical dynamical systems (see \cite{ES, ECM, JPR}). However, contrary to the classical case, we do not expect that 
the function
$$
\alpha\mapsto e_{\rho, t}(\alpha)= \log \tr (\rho \e^{t\cL_{(\alpha)}}(\one)),
$$
satisfies  this symmetry for fixed finite time $t$. A notable exception is provided by the very special 
"chaotic state"  $\rho=\rho_{\rm ch}=\one/\dim\cH$. Indeed, it follows from the fact that
$\Theta\circ\cL_{(\alpha)}^\ast=\cL_{(1-\alpha)}\circ\Theta$ (see the proof of Theorem~\ref{thm-1})
that $e_{\rho_{\rm ch},t}(1-\alpha)=e_{\rho_{\rm ch},t}(\alpha)$ for all $\alpha\in\rr^M$ and all $t\ge0$.\newline
{\bf Remark 2.} Property (8) is a consequence of energy conservation. It was first proposed by Andrieux {\sl et al.}\;in
the framework of Hamiltonian dynamics on the basis of a formal calculation \cite{AGMT}.

We shall call  the probability measure $P_\rho^t$ the {\sl entropic full counting statistics} (EFCS) of 
the QDS generated by $\cL$ (w.r.t.\;the specific decomposition $\cL=\sum_j\cL_j$). As explained in 
Section~\ref{sect-FCS}, in cases where this QDS arises as a weak coupling limit of the dynamics of 
a system ${\cal S}$ coupled to $M$ thermal reservoirs, the EFCS is the scaling limit of
a measure $\mathbb{P}_\rho^t$ which describes the mean rate of entropy exchange between 
the system and the $M$ reservoirs during the time interval $[0,t]$ (see Eq.~\eqref{FCS-limit} below).

An alternative  interpretation of  
the measures $P_\rho^t$ is based on the well-known unraveling technique. 
In other words, these measures can be understood in terms of a  classical stochastic process
which provides a coarse grained description of the dynamics of the system by so called quantum 
trajectories. Within this framework, $P_\rho^t$ is the joint distribution of $M$ random variables
which describe the exchange of entropy between the system and the $M$ reservoirs
(see Section~\ref{sec-unravel}).

\subsection{Entropic fluctuations}

As a direct consequence of Theorem~\ref{thm-1} and the G\"artner-Ellis theorem
(see, {\sl e.g.,} \cite{DZ,El}), we have

\begin{corollary}\label{prop-LDP} 
Assume that Hypotheses (DB) and (ER) hold  and let
$$
I(\varsigma)=-\inf_{\alpha\in\rr^M}\left(\alpha\cdot\varsigma+e(\alpha)\right).
$$
$I(\varsigma)$ is the Fenchel-Legendre transform of $e(-\alpha)$. Then:
\begin{enumerate}[{\rm (1)}]
\item $I(\varsigma)$ takes values in $[0, \infty]$ and is a convex lower-semicontinuous function with compact level sets.
\footnote{The level sets of $I$ are $\{\varsigma\,|\, I(\varsigma)\leq l\}$ where $l \in [0, \infty[$.}
\item $I(\varsigma)=0$  iff $\varsigma=\bar\varsigma$, where $\bar \varsigma=-\nabla e(0)$. 
Moreover, for any $\epsilon>0$ there exists a positive constant $a(\epsilon)$ such that
$$
P_\rho^t(\{\varsigma\in\rr^M\,|\,|\varsigma-\bar\varsigma|\ge\epsilon\})\le\e^{-t a(\epsilon)},
$$
for all $t>0$.
\item The family of measures $\{P_\rho^t\}_{t\ge0}$ satisfies the  large deviation principle 
with rate function $I$. More precisely, for any Borel set $G\subset\rr^M$ we have
\beq
-\inf_{\varsigma\in\mathrm{int}(G)} I(\varsigma) 
\leq\liminf_{t \rightarrow \infty} 
\frac{1}{t} \log P_\rho^t\left(G\right)
\leq\limsup_{t \rightarrow \infty} 
\frac{1}{t} \log P_\rho^t\left(G\right) \leq 
-\inf_{\varsigma\in\mathrm{cl}(G)} I(\varsigma),
\label{LDP-b}
\eeq
where $\mathrm{int}(G)$ and $\mathrm{cl}(G)$ denote the interior and the closure of the set $G$.
\item If Hypothesis {\rm (TR)} is satisfied, then the rate function satisfies 
\beq\label{es-snow1}
I(-\varsigma)=1\cdot\varsigma + I(\varsigma).
\eeq
\item If Hypothesis {\rm (\KMSB)} is satisfied, then $I(\varsigma)=+\infty$ for any $\varsigma\in\rr^M$ such that
$\beta^{-1}\cdot\varsigma\not=0$.
\end{enumerate}
\end{corollary}
{\bf Remark 1.} The  components
of $\bar\varsigma=(\bar\varsigma_1, \cdots, \bar \varsigma_M)$ describe the asymptotic rates of 
entropy transport between the system $\cS$ and the $M$ reservoirs constituting its environment. 
The non-negative number 
\[
\sigma_+=1\cdot\bar\varsigma=\sum_j \bar \varsigma_j,
\]
is the steady state {\em entropy production rate} of a QDS satisfying Hypotheses (ER) and (DB)
(see the next section for additional information about this important concept). If (TR) holds, 
then Relation \eqref{es-snow1} implies $I(-\bar \varsigma)=\sigma_+$ and $\sigma_+>0$ iff  
$\bar \varsigma \not=0$. \newline
{\bf Remark 2.} 
The large deviation principle (\ref{LDP-b})  quantifies the exponential rate of decay of the measures $P_\rho^t$ 
away from the asymptotic mean value $\bar\varsigma$ and describes the statistics of the fluctuations of the rates
of entropy transport over large but finite periods of time. In particular, (\ref{LDP-b}) implies that 
$$
P_\rho^t(\{\varsigma\in\rr^M\,|\,\varsigma \simeq\varphi\})\simeq\e^{-tI(\varphi)},
$$
for large $t$. Combining Parts (2) and (3) we derive that for large $t$, 
\beq\label{es-snow2}
\frac{P_\rho^t(\{\varsigma\in\rr^M\,|\,\varsigma\simeq-\varphi\})}{P_\rho^t(\{\varsigma\in\rr^M\,|\,\varsigma\simeq\varphi\})}
\simeq\e^{-t1\cdot\varphi}, 
\eeq
and in particular that 
\beq\label{es-snow3}
\frac{P_\rho^t(\{\varsigma\in\rr^M\,|\,\varsigma\simeq-\bar \varsigma\})}{P_\rho^t(\{\varsigma\in\rr^M\,|\,\varsigma\simeq\bar \varsigma\})}
\simeq\e^{-t \sigma_+}.
\eeq
The identities (\ref{es-snow}) and  (\ref{es-snow1}), together with the resulting asymptotics (\ref{es-snow2}) and 
(\ref{es-snow3}), constitute {\em fluctuation relations} for a  QDS out of equilibrium. One important feature of 
the fluctuation relations is universality (independence of the model). \newline

Theorem~\ref{thm-1} and Bryc's theorem (see Proposition~1 in \cite{Br} and Appendix A in \cite{JOPP}) imply the 
Central Limit Theorem for the family of measures $\{P_\rho^t\}_{t\geq 0}$. 
\begin{corollary}\label{prop-CLT}
Assume that Hypotheses (ER) and (DB) hold. Then for any Borel set $G\subset \rr^M$, 
\begin{equation}\label{clt-eq}
\lim_{t\to\infty}P_\rho^t\left(
\left\{\varsigma\in\rr^M\,\bigl|\,\sqrt{t}(\varsigma-\langle\varsigma\rangle_{\rho,t})\in G\right\}\right)=\mu_D(G),
\end{equation}
where $\mu_D$ denotes the centered Gaussian measure on $\rr^M$ with covariance $D$ given by
$$
D_{ij}=\left.\frac{\partial^2e(\alpha)}{\partial\alpha_i\partial\alpha_j}\right|_{\alpha=0}.
$$
\end{corollary}

Note that if  Hypothesis (\KMSB) holds, then Theorem~\ref{thm-1}~(8) implies  that the Gaussian
measure $\mu_D$ has its support on the hyperplane $\beta^{-1}\cdot\varsigma=0$. This is of course
related to Part (5) of Corollary~\ref{prop-LDP} and to energy conservation.

%%%%%%%%%%%%%%%%%
\subsection{Thermodynamics}
%%%%%%%%%%%%%%%%%
The von Neumann entropy of a state $\rho$ is $\Ent(\rho)=-\tr(\rho\log\rho)$ and we shall call
$S=-\log\rho$  the entropy observable associated to $\rho$. The relative entropy of a state $\nu$ 
w.r.t.\;to another state $\mu$ is
$$
\Ent(\nu|\mu)=\left\{
\begin{array}{ll}
\tr(\nu(\log\mu-\log\nu))&\text{if }{\rm Ran}(\nu)\subset{\rm Ran}(\mu);\\[10pt]
-\infty&\text{otherwise}.
\end{array}
\right.
$$
We refer the reader to the monograph of Ohya and Petz \cite{OP} for further information on these
fundamental concepts.  Following Lebowitz and Spohn \cite{LS1, Sp3}, we define the entropy production in the state $\rho$
of a QDS $\{\e^{t\cL}\}_{t\ge0}$ satisfying Hypothesis~(DB) by \footnote{The derivative exists for all $\rho \in \fS$, see Theorem~3 in 
\cite{Sp3}.}
\beq\label{def-entp}
\sigma(\rho)=\frac{\d\ }{\d t}\sum_{j=1}^M\Ent(\e^{t\cL_j^\ast}(\rho)|\rho_j)\Big|_{t=0}.
\eeq
We recall basic properties of the entropy production established in  \cite{ LS1, Sp3}.
\begin{enumerate}[{\rm (a)}] 
\item Since $\rho_j$ is a steady state of the QDS generated by $\cL_j$, we have 
$$
\Ent(\e^{t\cL_j^\ast}(\rho)|\rho_j)=\Ent(\e^{t\cL_j^\ast}(\rho)|\e^{t\cL_j^\ast}(\rho_j)),
$$
and Uhlman's monotonicity theorem (\cite{Uh}, see also \cite{OP, JOPP}) implies that the r.h.s.\;of this 
identity is a non-decreasing function of $t$. Hence, 
\[ \sigma(\rho)\geq 0.\]
\item  An application of a theorem of Lieb \cite{Lb} gives that the map ${\mathfrak S} \ni \rho \mapsto \sigma(\rho)$ is convex (see 
Theorem~3 in \cite{Sp3}).
\item Set $S_j=-\log \rho_j$ and $\cI_j=\cL_j(S_j)$. An immediate consequence of  (\ref{def-entp})   is the {\em entropy balance equation}:
\beq
\frac{\d\ }{\d t}\Ent(\e^{t\cL^\ast}(\rho))\Big|_{t=0}=\sigma(\rho)
+\sum_{j=1}^M\rho(\cI_j).
\label{eq-ebal}
\eeq
The second term on the r.h.s.\;of Eq.~\eqref{eq-ebal} describes the flux of entropy entering the system. 
Thus, we can interpret the observable $\cI_j$ as the entropy flux out of the
$j^{\rm th}$ reservoir. Note that if $\rho$ is a steady state, then the l.h.s.\;of (\ref{eq-ebal}) vanishes, and the 
entropy balance equation takes the form 
\begin{equation}\label{airport-1}
\sigma(\rho)=-\sum_{j=1}^M\rho(\cI_j).
\end{equation}
\end{enumerate}

Our next result links the function $e(\alpha)$ to the observables
$S_j$ and $\cI_j$.

\bet\label{prop-Flux}
Let $\{\e^{t\cL}\}_{t\ge0}$ be a QDS satisfying Hypotheses (ER) and (DB). Set $\cJ_j=\cI_j-\rho_+(\cI_j)$. 
Then the following holds:
\begin{enumerate}[{\rm (1)}]
\item 
$$
\left.\frac{\partial e(\alpha)}{\partial\alpha_j}\right|_{\alpha=0}=\rho_+(\cI_j).
$$
In particular, 
\[
\rho_+(\cI_j)=-\lim_{t\to\infty}\langle\varsigma_j\rangle_{\rho,t}=-\bar\varsigma_j,
\]
and $\sigma(\rho_+)=\sum_j \bar \varsigma_j$.
\item 
\begin{align*}
\left.\frac{\partial^2 e(\alpha)}{\partial\alpha_j\partial\alpha_k}\right|_{\alpha=0}
&=-\int_0^\infty\rho_+\left(
\e^{t\cL}(\cJ_j)\cJ_k^++\e^{t\cL}(\cJ_k)\cJ_j^+\right)\d t\\
&+\int_0^\infty\rho_+\left(\cL_k(\e^{t\cL}(\cJ_j)S_k)+\cL_j(\e^{t\cL}(\cJ_k)S_j)\right)\d t
+\delta_{jk}\rho_+(D_j(S_j,S_j))\\[4pt]
&=\lim_{t\to\infty}t\langle(\varsigma_j-\langle\varsigma_j\rangle_{\rho,t})(\varsigma_k-\langle\varsigma_k\rangle_{\rho,t})\rangle_{\rho,t},
\end{align*}
where $\cJ_j^+=\cL_j^{\rho_+}(S_j)=\cL_j^\ast(S_j\rho_+)\rho_+^{-1}$ and
$D_j(A,B)=\cL_j(A^\ast B)-\cL_j(A^\ast)B-A^\ast\cL_j(B)$ is the dissipation function of the $j^{\rm th}$ Lindbladian.
\end{enumerate}
\eet
{\bf Remark.} Under the assumptions of the theorem the semigroup $\semigroup$ is relaxing exponentially fast to 
$\rho_+$. Since $\rho_+(\cJ_j)=0$, this implies that the operators $\e^{t\cL}(\cJ_j)$ are exponentially
decaying as $t\to\infty$, and so the time integrals in Part (2) are absolutely convergent. \newline
{\bf Remark 2.}  We shall make use of Part (2) in Section \ref{Sect-LRT} where we discuss linear response theory. 

%%%%%%%%%%%%%%%%%%%%%%
\subsection{Energy fluxes}\label{Sect-EF}
%%%%%%%%%%%%%%%%%%%%%%%
The Hypothesis (\KMSB) allows us to relate entropy fluxes to energy fluxes by simple rescaling and to 
restate our main results in terms of energy transport. As a preparation for the discussion of the linear response 
theory, in this section we briefly discuss how this restating is carried out. Until the end of this section we shall assume that 
Hypotheses (ER), (DB), and (\KMSB) hold. 

The observable describing the energy flux  out of the  $j^{\rm th}$  reservoir is 
${\cal F}_j=\cL_j(H_\cS)$ (see \cite{LS1}). Note that ${\cal I}_j=\beta_j {\cal F}_j$. If in addition (TR)
holds, then $\Theta(H_\cS)=H_\cS$ and it follows from Parts (1) and (2) of Theorem~\ref{thm-DB} that 
$\cL_j^{\rho_j}(H_\cS)= \cL_j(H_\cS)$. Hence, 
\[\Theta({\cal F}_j)=\cL_j^{\rho_j}(\Theta(H_\cS))={\cal F}_j.\]
The steady state energy fluxes are 
\[
\bar \phi_j=\rho_+({\cal F}_j).
\]
Obviously, $\bar \varsigma_j =-\beta_j \bar \phi_j$, and Eq. (\ref{airport-1})   takes the form 
\beq
\sigma(\rho_+)=-\sum_{j=1}^M \beta_j\bar \phi_j\geq 0.
\label{seclaw}
\eeq
This relation expresses  the second law of thermodynamics for QDS satisfying our assumptions. 
The relation $\cL^\ast(\rho_+)=0$ yields the first law (conservation of energy):
\beq
\sum_{j=1}^M \bar \phi_j=0.
\label{firstlaw}
\eeq
The energetic full counting statistics of the system is the probability measure $Q_{\rho}^t$ on 
$\rr^M$ given by 
\[ 
Q_{\rho}^t(\phi)= P_\rho^t(-\beta\phi), 
\]
where $\beta \phi=(\beta_1\phi_1, \cdots, \beta_M\phi_M)$. In particular, 
\[
\tr (\rho\e^{t\cL_{(\alpha/\beta)}}(\one))=\int_{\rr^M}\e^{t\alpha\cdot\phi}\d Q_{\rho}^t(\phi),
\]
where $\alpha/\beta=(\alpha_1/\beta_1, \cdots, \alpha_M/\beta_M)$. Hence, for $\alpha \in \rr^M$,
\[
\chi(\alpha)=\lim_{t \rightarrow \infty}\frac{1}{t}\log 
\int_{\rr^M}\e^{t \alpha \cdot \phi}\d Q_{\rho}^t(\phi)=e(-\alpha/\beta),
\]
and in particular, 
\beq
\left.\frac{\partial \chi (\alpha)}{\partial \alpha_j}\right|_{\alpha =0}=\bar \phi_j, \qquad \left.\frac{\partial^2\chi(\alpha)}{\partial \alpha_j\partial \alpha_k}
\right|_{\alpha=0}= \frac{1}{\beta_j\beta_k}\left.\frac{\partial^2e(\alpha)}{\partial \alpha_j\partial \alpha_k}
\right|_{\alpha=0}.
\label{ef-1}
\eeq
Note that the translation symmetry of $e(\alpha)$ (described in Part~(8) of Theorem~\ref{thm-1}) implies 
that  
\beq
\chi(\alpha)=\chi(\alpha+\lambda 1),
\label{ef-2}
\eeq
for all $\alpha \in \rr^M$, $\lambda \in \rr$. If (TRI) holds, then the Evans-Searles symmetry takes the form 
\beq
\chi(\alpha)=\chi(-\beta-\alpha).
\label{ef-3}
\eeq
The large $t$ fluctuations of $Q_\rho^t$ are described by obvious reformulations of Corollaries \ref{prop-LDP} and 
\ref{prop-CLT}.

Finally, we discuss briefly the equilibrium case where $\beta_j=\beta_0$ for $j=1, \cdots, M$.
In this case
\[
\rho_j=\rho_0=\frac{\e^{-\beta_0 H_\cS}}{\tr (\e^{-\beta_0 H_\cS})},
\]
and $\cL_j^\ast(\rho_0)=0$ for all $j$. It follows that $\cL^\ast(\rho_0)=0$ and hence that 
$\rho_+=\rho_0$ and $\bar \phi_j=0$ for all $j$. Combining Parts (1) and (2) of 
Theorem~\ref{thm-DB} with Theorem~\ref{prop-Flux} (2) one easily derives that 
${\cal J}_j^+= {\cal J}_j=\mathcal{I}_j=\beta_0{\cal F}_j$, and that 
\begin{equation}\label{ef-4}
\left.\frac{\partial^2\chi(\alpha)}{\partial \alpha_j\partial \alpha_k}\right|_{\alpha=0}
=-\int_0^\infty\rho_0\left(\e^{t \cL}({\cal F}_j){\cal F}_k + \e^{t \cL}({\cal F}_k){\cal F}_j\right)\d t
+\delta_{jk}\rho_0(D_j(H_\cS, H_\cS)).
\end{equation}
If the pair $(\rho_0, \cL)$ is TRI, then 
\beq\label{snow}
\rho_0(\e^{t \cL}({\cal F}_j){\cal F}_k)= 
\rho_0(\Theta({\cal F}_k\e^{t \cL}({\cal F}_j)))=
\rho_0({\cal F}_k\e^{t \cL^{\rho_0}}({\cal F}_j)) =
\rho_0(\e^{t\cL}({\cal F}_k){\cal F}_j).
\eeq
%%%%%%%%%%%%%%%%%%%%%%
\subsection{Linear response theory}\label{Sect-LRT}
%%%%%%%%%%%%%%

Our last result concerns linear response to thermodynamical forces. We consider a small system
$\cS$ coupled to $M$ thermal reservoirs $\cR_j$ in equilibrium at inverse temperatures
$\beta_j$ where each $\beta_j$ is close to some common equilibrium value $\beta_0>0$.
The purpose of linear response theory is to study the behavior of various physical quantities
to first order in the thermodynamical forces $\zeta_j=\beta_0-\beta_j$. It is therefore natural to 
parametrize $\beta=(\beta_1,\ldots,\beta_M)$ by $\zeta=(\zeta_1, \cdots, \zeta_M)$ so that 
$\zeta=0$ corresponds to the equilibrium situation
$\beta=\beta_\eq=(\beta_0,\ldots,\beta_0)$. The precise setup is as follows.

Let $(\cL_\zeta)_{\zeta\in {\cal U}}$ be a family of Lindbladians indexed by an open
neighborhood $\cal U$ of $0$ in $\rr^M$ and such that
each $\cL_\zeta$ satisfies Hypotheses (ER) and (TR). Moreover, we assume Hypotheses
(DB) and (\KMSB) in the following form: for each $\zeta\in{\cal U}$,
$$
\cL_\zeta= \sum_{j=1}^M \cL_{\zeta,j},
$$ 
where $\cL_{\zeta,j}$ depends only on $\zeta_j$ and satisfies the detailed balance condition
w.r.t.\;the state 
$$
\rho_{\zeta_j}=\frac{\e^{-(\beta_0-\zeta_j) H_\cS}}{\tr(\e^{-(\beta_0-\zeta_j)H_\cS})},
$$
for some $\zeta$-independent self-adjoint $H_\cS\in\cO$. We shall also assume the following 
regularity in $\zeta$:
\begin{quote} {\bf (RE)} The map $ \zeta \mapsto \cL_{\zeta}$ is continuously 
differentiable at $\zeta=0$.
\end{quote}
In what follows we shall indicate explicitly the dependence on $\zeta$ by writing $\cL_{\zeta, (\alpha)}$,
$e(\zeta,\alpha)$, $\chi(\zeta,\alpha)$, $\mathcal{F}_{\zeta,j}$, $\bar\phi_{\zeta, j}$, etc.
Our assumptions imply that all partial derivatives of $\cL_{\zeta, (\alpha)}$ w.r.t.\;$\alpha$
are continuously differentiable w.r.t.\;$\zeta$ at $\zeta=0$.

For all $\alpha\in\rr^M$ and $\zeta\in\cal U$, $e(\zeta,\alpha)$ is a simple 
eigenvalue of $\cL_{\zeta,(\alpha)}$. The perturbation theory of isolated eigenvalues 
(see the proof of Theorem~\ref{prop-Flux}) implies that all partial derivatives of $e(\zeta,\alpha)$
w.r.t.\;$\alpha$ are also continuously differentiable w.r.t.\;$\zeta$ at $\zeta=0$
and the same holds for the function $\chi(\zeta,\alpha)$. In particular, the maps 
$\zeta\mapsto\bar\phi_{\zeta, j}$ are continuously differentiable at $\zeta=0$.

Combining \eqref{seclaw} and \eqref{firstlaw} yields the following expressions of
the first and second laws of thermodynamics
\[
\sum_{j=1}^M \bar \phi_{\zeta, j}=0, \qquad \sum_{j=1}^M \zeta_j\bar \phi_{\zeta, j}\geq 0.
\]
The kinetic transport coefficients are defined by 
\[
L_{jk}=\frac{\partial \bar \phi_{\zeta, j}}{\partial \zeta_k}\Big|_{\zeta=0}.
\] 
It follows from the first law that 
\beq
\sum_{j=1}^ML_{jk}=0,
\label{con}
\eeq
while the second law implies that the {\em real} quadratic form determined by the matrix
$[L_{jk}]$ is positive definite.\footnote{This does not imply that $L_{jk}=L_{kj}$.}
It further follows from the first relation in \eqref{ef-1} that
\[
L_{jk}=\left.\frac{\partial^2\chi (\zeta, \alpha)}{\partial \zeta_k \partial \alpha_j}\right|_{\zeta=\alpha=0}.
\]
In terms of the variable $\zeta$, the Evans-Searles symmetry \eqref{ef-3} takes the form 
$\chi(\zeta,\alpha)=\chi(\zeta, -\beta_{\eq} +\zeta-\alpha)$, while the translation symmetry \eqref{ef-2} 
reads $\chi(\zeta, \alpha)=\chi(\zeta, \alpha +\lambda 1)$. Since $\beta_\eq=\beta_0 1$, combining 
these two symmetries we derive 
\begin{equation}\label{lrt-1}
\chi (\zeta, \alpha)=\chi(\zeta, \zeta-\alpha).
\end{equation}
This relation and the chain rule (see Lemma~4.4 in \cite{JPR}) yield 
\begin{equation}
L_{jk}=\left.\frac{\partial^2\chi (\zeta, \alpha)}{\partial \zeta_k \partial \alpha_j}\right|_{\zeta=\alpha=0}
=-\frac{1}{2}\left.\frac{\partial^2\chi (\zeta, \alpha)}{\partial \alpha_k \partial \alpha_j}
\right|_{\zeta=\alpha=0}.
\label{lrt-2}
\end{equation}
The equality of mixed partial derivatives 
$\partial_{\alpha_k}\partial_{\alpha_j}\chi=\partial_{\alpha_j}\partial_{\alpha_k}\chi$ 
implies the Onsager reciprocity relations 
$L_{jk}=L_{kj}$. Relations (\ref{ef-4}), (\ref{snow}),  and Corollary~\ref{prop-CLT}  complete the linear 
response theory. We summarize:

\bet\label{thm-LRT} Under the Hypotheses formulated at the beginning of this section the following
statements hold.
\begin{enumerate}[{\rm (1)}] 
\item The Green-Kubo formulae:
\[L_{jk}=\int_0^\infty \rho_0(\e^{t \cL_0}({\cal F}_{0, j}){\cal F}_{0, k})\d t - \frac{1}{2}
\delta_{jk}\rho_0(D_{0,j}(H_\cS, H_\cS)),
\]
where $D_{0,j}$ denotes the dissipation function of $\cL_{0,j}$.
\item The Onsager reciprocity relations:
\[ L_{jk}=L_{kj}.\]
\item The Fluctuation-Dissipation Theorem: for a state $\rho$ on $\cO$ let $Q_{\eq,\rho}^t$ be the 
energetic full counting statistics of the equilibrium system, {\sl i.e.,}
$$
\tr(\rho\e^{t\cL_{0,(\alpha/\beta_0)}}(\one))=\int_{\rr^M}\e^{t\alpha\cdot\phi}\d Q^t_{\eq,\rho}(\phi).
$$
and let $\langle\,\cdot\,\rangle_{\eq,\rho,t}$ denote the expectation w.r.t.\;the measure 
$Q_{\eq,\rho}^t$. For any Borel set
$G\subset \rr^M$,
\[
\lim_{t\rightarrow \infty} Q_{\eq,\rho}^t\left(\left\{ 
\phi \in \rr^M\, \big|\, \sqrt t(\phi-\langle \phi\rangle_{\eq,\rho,t})\in G\right\}\right)=\mu_D(G),
\]
where $\mu_D$ is the centered Gaussian measure on $\rr^M$ with covariance $D$ given by 
\[
D_{jk}=2L_{jk}.
\]
\end{enumerate}
\eet
{\bf Remark 1.} Concerning the diagonal transport coefficients $L_{jj}$, the terms 
$\rho_0(D_{0,j}(H_\cS, H_\cS))$ are non-negative, and are strictly positive if $\cS$ is effectively 
coupled to the $j^{\rm th}$-reservoir (see Section \ref{sect-WCL}). Parts (1)-(2) of 
Theorem~\ref{thm-DB} imply that 
$\rho_0(D_{0,j}(H_\cS, H_\cS))=-2\rho_0(H_\cS{\cal F}_{0,j})$.\newline
{\bf Remark 2.} In the absence of time-reversal, Part (3) holds with 
\[
D_{jk}=\int_0^\infty\rho_0(\e^{t \cL_0}({\cal F}_{0, j}){\cal F}_{0, k}
+\e^{t \cL_0}({\cal F}_{0, k}){\cal F}_{0, j})\d t
-\delta_{jk}\rho_0(D_{0,j}(H_\cS, H_\cS)).
\]
{\bf Remark 3.} Parts (1) and (2) of Theorem~\ref{thm-LRT} were first proven in \cite{LS1} by a 
different method. For comparison purposes we sketch the proof of \cite{LS1}. 
Since $\cL_{\zeta,k}^\ast(\rho_{\zeta_k})=0$,
\[
\label{lesp-1}
\frac{\d\cL_{\zeta, k}^\ast}{\d\zeta_k}(\rho_{\zeta_k})
=-\cL_{\zeta, k}^\ast\left(\frac{\d\rho_{\zeta_k}}{\d\zeta_k}\right)
=\cL_{\zeta, k}^\ast(H_\cS\rho_{\zeta_k})={\cal F}_{\zeta, k}\rho_{\zeta_k},
\]
where  the last equality follows from Parts (1) and (2) of Theorem~\ref{thm-DB}. 
Hypotheses (ER) and (RE) imply that the map $\zeta\mapsto \rho_{\zeta, +}$ is continuously 
differentiable at $\zeta=0$. Differentiating $\cL_{\zeta}^\ast(\rho_{\zeta, +})=0$ w.r.t.\;$\zeta_k$
at $\zeta=0$, we get  
\[
\label{lesp-2}
\frac{\d\cL_{\zeta, k}^\ast}{\d\zeta_k}\Big|_{\zeta=0}(\rho_0)
=-\cL_0^\ast\left(\frac{\partial \rho_{\zeta,+}}{\partial\zeta_k}\Big|_{\zeta=0}\right).
\]
The last two relations give  
\beq
\label{lesp-3}
\cL_0^\ast\left(\frac{\partial\rho_{\zeta,+}}{\partial\zeta_k}
\Big|_{\zeta=0}\right)=-{\cal F}_{0, k}\rho_0.
\eeq
Since 
\beq
\lim_{t\rightarrow\infty}\e^{t \cL_0^\ast}({\cal F}_{0,k}\rho_0)=\rho_0({\cal F}_{0, k})\rho_0=0,
\label{Lstarlim}
\eeq
the operators $\e^{t\cL_0^\ast}({\cal F}_{0, k}\rho_0)$ are exponentially decaying as $t\rightarrow\infty$,
and we  deduce from  (\ref{lesp-3}) that there is a constant $c$ such that
\[
\frac{\partial\rho_{\zeta,+}}{\partial\zeta_k}\Big|_{\zeta=0}
=c\rho_0+\int_0^\infty\e^{t\cL_0^\ast}({\cal F}_{0,k}\rho_0)\d t.
\]
If $j\not=k$ then $\mathcal{F}_{\zeta,j}=\cL_{\zeta,j}(H_\cS)$ does not depend on $\zeta_k$ and
it follows that
\[
L_{jk}=\tr\left({\cal F}_{0,j}\frac{\partial\rho_{\zeta,+}}{\partial\zeta_k}\Big|_{\zeta=0}\right)
=\int_{0}^{\infty}\rho_0(\e^{t \cL_0}({\cal F}_{0,j}){\cal F}_{0, k})\d t.
\]
The conservation law \eqref{con}, the limit \eqref{Lstarlim} and the last formula in Remark 1 yield 
\begin{align*}
L_{kk}=-\sum_{j:j\not=k}L_{jk}&=\int_0^\infty
\rho_0(\e^{t \cL_0}({\cal F}_{0, k}){\cal F}_{0,k}-\e^{t \cL_0}(\cL_0(H_\cS)){\cal F}_{0,k})\d t\\
&=\int_0^\infty\rho_0(\e^{t \cL_0}({\cal F}_{0, k}){\cal F}_{0,k})\d t
-\int_0^\infty\frac{\d\ }{\d t}\tr(\e^{t \cL_0^\ast}({\cal F}_{0,k}\rho_0)H_\cS)\d t\\
&=\int_0^\infty\rho_0(\e^{t \cL_0}({\cal F}_{0, k}){\cal F}_{0,k})\d t
+\rho_0(H_\cS{\cal F}_{0,k})\\
&=\int_0^\infty \rho_0(\e^{t \cL_0}({\cal F}_{0,k}){\cal F}_{0,k})\d t
-\frac{1}{2}\rho_0(D_{0,k}(H_\cS, H_\cS)).
\end{align*}
Note that the  above argument did not make use of Hypothesis (TR) and so Part (1) of Theorem~\ref{thm-LRT} holds without time-reversal 
assumption (in  fact, Lebowitz and Spohn do not discuss time-reversal at all in \cite{LS1}). However,  if 
 the pair $(\rho_0, \cL_0)$ is time-reversal invariant, then Part (1) and Relation \eqref{snow} yield the 
Onsager reciprocity relations. 

In contrast to the direct argument of \cite{LS1}, the  proof  described in this section exploits fundamentally the symmetry (\ref{lrt-1}).   The  advantage of this 
derivation in context of a QDS out of equilibrium is conceptual. The fluctuation
relations are structural model independent features of non-equilibrium statistical mechanics. As observed by 
Gallavotti \cite{Ga}, in the linear regime 
near equilibrium the fluctuation relations  reduce to familiar fluctuation-dissipation formulae, and this  structural  model independent view of  linear response 
theory is of fundamental conceptual importance  (see \cite{LS2, JPR, JOPP} for a 
pedagogical discussion of this point). Our proof shows how 
a QDS out of equilibrium fit into this general picture and complements the derivation of \cite{LS1} from the conceptual 
point of view.
%%%%%%%%%%%%%%%%%%%%%%%%%%%%%%%%%%%%%%%%%%%%%%%
%%%%%%%%%%%%%%%%%%%%%%%%%%%%%%%%%%%%%%%%%%%%%%%
\section{Weakly coupled open quantum systems}
\label{sect-WCL}

We consider a small quantum system $\cS$, described by the Hamiltonian $H_\cS$
acting on the finite dimensional Hilbert space $\cH_\cS$. To induce a dissipative dynamics
on $\cS$, we couple this system to several infinitely extended thermal reservoirs $\cR_1,\ldots,\cR_M$. 
Each reservoir $\cR_j$ is initially  in a thermal equilibrium state at inverse temperature 
$\beta_j>0$.\footnote{Here, we could also consider conserved charges and introduce associated
chemical potentials. We refrain to do so in order to keep notation as simple as possible.}
By passing to the GNS representations induced by these states, each  $\cR_j$ is described by a Hilbert 
space $\cH_j$, a $W^\ast$-algebra $\cO_j\subset\cB(\cH_j)$ of observables, and a self-adjoint operator
$L_j$ (the Liouvillean) acting on $\cH_j$, such that the Heisenberg dynamics $\tau_j^t(A)=\e^{\i tL_j}A\e^{-\i tL_j}$
leaves $\cO_j$ invariant. The initial state of $\cR_j$ is given by
$\cO_j\ni A\mapsto\omega_j(A)=\langle\xi_j|A\xi_j\rangle$,  where 
$\xi_j\in\cH_j$ is a unit vector such that $L_j\xi_j=0$. Moreover, the state $\omega_j$ satisfies  the KMS boundary condition: for all 
$A, B \in \cO_j$, 
\beq
\omega_j(A\tau_j^{t}(B))=\omega_j(\tau_j^{t-\i\beta_j}(B)A).
\label{KMScond}
\eeq

The Hilbert space of the joint system $\cS+\cR_1+\cdots+\cR_M$ is 
$\cH=\cH_\cS\otimes\cH_1\otimes\cdots\otimes\cH_M$ and we shall denote
$H_\cS\otimes \one\otimes\cdots\otimes \one$, $\one\otimes H_1\otimes\cdots\otimes \one$, \ldots\,
simply by $H_\cS$, $H_1$,\ldots

The interaction between the system $\cS$ and the reservoir $\cR_j$ is described  by the Hamiltonian
$$
H_{\cS\cR_j}=\sum_{k=1}^{n_j}Q^{(k)}_j\otimes R^{(k)}_j,
$$ 
where each $Q^{(k)}_j$ is a  self-adjoint operator on 
$\cH_\cS$ and each $R^{(k)}_j$ is a  self-adjoint element of $\cO_j$ such that $\omega_j(R^{(k)}_j)=0$.\footnote{In some models 
(like the spin-boson system) the operators $R^{(k)}_j$ are unbounded and only affiliated to  the $W^\ast$-algebra ${\cal O}_j$. 
With some additional technicalities the discussions  of this and the next three section easily extend to such cases, see any of the references 
\cite{DF, DJP, dR2, JPPW, LS1}.}
The full Hamiltonian (more precisely the semi-standard Liouvillean in the terminology of \cite{DJP})
of the coupled system is
$$
L_\lambda=H_\cS+\sum_{j=1}^M\left(L_j+\lambda H_{\cS\cR_j}\right),
$$
where $\lambda$ is a coupling constant. The effective dynamics of the system $\cS$
is then defined by the family of linear map $\{\cT_\lambda^t\}_{t\in\rr}$ on $\cB(\cH_\cS)$
determined by
$$
\langle\psi|\cT_\lambda^t(X)\psi\rangle
=\langle\psi\otimes\xi|\e^{\i tL_\lambda}(X\otimes \one)\e^{-\i tL_\lambda}\psi\otimes\xi\rangle,
$$
where  $X\in\cB(\cH_\cS)$, $\psi \in\cH_\cS$, and  
$\xi=\xi_1\otimes\cdots\otimes\xi_M$.

Except in trivial cases, $\{\cT_\lambda^t\}_{t\geq 0}$ is not a semigroup.  However,
under appropriate conditions on the decay of the multi-time correlation functions
$\omega_j(\tau_j^{t_1}(R^{(k_1)}_j)\cdots\tau_j^{t_n}(R^{(k_n)}_j))$, Davies has shown 
(see Theorem~2.3 in \cite{Da1}) that there exists a Lindbladian ${\cal L}$ generating a QDS 
such that $\cL$  commutes with $\cL_\cS(X)=\i[H_\cS,X]$, and
$$
\lim_{\lambda\to0}\sup_{\lambda^2t\in I}\|\cT_\lambda^{t}-\e^{t(\cL_\cS+\lambda^2\cL)}\|=0,
$$
holds for any compact interval $I=[0,\tau]\subset\rr$. In other words, in the limit of small coupling $\lambda\to0$ and for times of the order $\lambda^{-2}$
the effective dynamics of $\cS$ is well approximated by the quantum dynamical semigroup
generated by $\cL_\cS+\lambda^2\cL$. This theory is well-known and we refer the reader 
to the in depth exposition of \cite{LS1, DF} for further details. 
To write down the explicit form of the generator $\cL$, we introduce the functions
$$
h_j^{(kl)}(\omega)
=\int_{-\infty}^\infty e^{-i\omega t}\langle\xi_j|R^{(k)}_j\tau_j^t(R^{(l)}_j) \xi_j\rangle\,\d t
=2\pi\langle R^{(k)}_j\xi_j|\delta(L_j-\omega)R^{(l)}_j\xi_j\rangle,
$$
and their Hilbert transforms
$$
s_j^{(kl)}(\omega)={\cal P}\!\!\int_{-\infty}^\infty 
\frac{h_j^{(kl)}(\nu)}{\nu-\omega}\,\frac{\d\nu}{2\pi}
=\langle R^{(k)}_j\xi_j|{\cal P}(L_j-\omega)^{-1}R^{(l)}_j\xi_j\rangle,
$$
where ${\cal P}$ denotes Cauchy's principal value (the hypotheses of the above mentioned
theorem of Davies ensure the existence of these integrals). Note that the $n_j\times n_j$-matrices
\[h_j(\omega)=[h_j^{(kl)}(\omega)],\qquad s_j(\omega)=[s_j^{(kl)}(\omega)],\]
are respectively
positive and self-adjoint and that the KMS condition (\ref{KMScond}) implies the relation
\begin{equation}\label{wco-1}
h_j^{(kl)}(-\omega)=\e^{-\beta_j\omega}h_j^{(lk)}(\omega).
\end{equation}
We denote by $P_\mu$ the spectral projection of $H_\cS$ associated to the 
eigenvalue $\mu\in\sp(H_\cS)$, and for 
\[\omega\in\Omega=\{\mu-\nu\,|\,\mu,\nu\in\sp(H_\cS)\},\] 
we define
\beq
V_j^{(k)}(\omega)= \sum_{\mu-\nu=\omega} P_\nu Q_j^{(k)} P_\mu = V_j^{(k)\ast}(-\omega).
\label{KrausFam}
\eeq
Obviously, 
\beq\label{KrausFam-1}
\e^{\alpha H_\cS}V_j^{(k)}(\omega)\e^{-\alpha H_\cS}=\e^{-\alpha \omega}V_j^{(k)}(\omega),
\eeq
for all $\alpha\in \cc$.

The generator $\cL$ has the Lindblad form (\ref{prelim-5}), with the self-adjoint 
operator $T$ given by
$$
T=\sum_{j=1}^MT_j,\qquad
T_j= \sum_{k,l=1}^{n_j}\sum_{\omega\in\Omega}s_j^{(kl)}(\omega)
V_j^{(k)\ast}(\omega) V_j^{(l)}(\omega),
$$
and the CP map $\Phi$ given by
$$
\Phi(X)=\sum_{j=1}^M\Phi_j(X),\qquad
\Phi_j(X)=\sum_{k,l=1}^{n_j}\sum_{\omega\in\Omega}h_j^{(kl)}(\omega)
V_j^{(k)\ast}(\omega)XV_j^{(l)}(\omega).
$$
A Kraus decomposition of $\Phi_j$ is constructed as follows. Denote by 
$u_j(\omega)=[u_j^{(kl)}(\omega)]$ a unitary matrix which diagonalize the positive matrix
$h_j(\omega)$, 
\[u_j(\omega)^\ast h_j(\omega)u_j(\omega)=[\delta_{kl}g_j^{(k)}(\omega)].\]
Setting $W_j^{(k)}(\omega)=\sqrt{g_j^{(k)}(\omega)}\sum_lu_j^{(kl)}(\omega)V_j^{(l)}(\omega)$, we
obtain
$$
\Phi_j(X)=\sum_{k=1}^{n_j}\sum_{\omega\in\Omega}W_j^{(k)\ast}(\omega)XW_j^{(k)}(\omega).
$$
Note that $\cL$ can be written as the sum of the Lindbladians
$$
\cL_j(X)=\i[T_j,X]-\frac12\{\Phi_j(\one),X\}+\Phi_j(X),
$$
where $\cL_j$ describes the interaction of the small system $\cS$ with a single reservoir $\cR_j$. Using (\ref{wco-1}) and 
(\ref{KrausFam-1}) 
one easily verifies  that $\cL_j$ satisfies the detailed balance condition w.r.t.\;the faithful state
\beq
\rho_j=\frac{\e^{-\beta_j H_\cS}}{\tr(\e^{-\beta_j H_\cS})}.
\label{rho-j-eq}
\eeq
Thus, Hypotheses (DB) and (\KMSB) are automatically satisfied by the weak coupling Lindbladian $\cL$.

Regarding time-reversibility, assuming that
\begin{enumerate}[{\rm (a)}]
\item each reservoir is time-reversal invariant, {\sl i.e.,} there exists
antiunitary involution $\theta_j$ acting on $\cH_j$ such that $L_j\theta_j=\theta_jL_j$
and $\theta_j\xi_j=\xi_j$;
\item the small system $\cS$ is time-reversal invariant, {\sl i.e.,} there is an antiunitary 
involution $\theta_\cS$ on $\cH_\cS$ such that  $\theta_\cS H_\cS=H_\cS\theta_\cS$;
\item $\theta_jR_j^{(k)}=R_j^{(k)}\theta_j$ and 
$\theta_\cS Q_j^{(k)}=Q_j^{(k)}\theta_\cS$ for all $j,k$, \end{enumerate}
we easily conclude that $
h_j^{(kl)}(\omega)=\overline{h_j^{(lk)}(\omega)}$, $s_j^{(kl)}(\omega)=\overline{s_j^{(lk)}(\omega)}$,
and 
$\theta_\cS V_j^{(k)}(\omega)=V_j^{(k)}(\omega)\theta_\cS$. It immediately follows that
$\theta_\cS T_j=T_j\theta_\cS$ and $\Phi_j(\theta_\cS X\theta_\cS)=\theta_\cS\Phi_j(X)\theta_\cS$.
Hence, 
Hypothesis (TR) is satisfied with $\Theta(X)=\theta_\cS X\theta_\cS$.

We now turn to   the ergodicity Hypothesis (ER).  Clearly, $\{Q_j^{(k)}\}_{j,k}^\prime\cap \{H_\cS\}^\prime \subset 
\Ker \cL$ and the condition 
\begin{equation}
\{Q_j^{(k)}\}_{j,k}^\prime\cap \{H_\cS\}^\prime =\cc \one,
\label{spohn-cond}
\end{equation}
is obviously necessary for (ER) to hold. On the other hand,
assuming that the matrices 
$h_j(\omega)$ are strictly positive for all $1\leq j \leq M$ and $\omega \in\Omega$, the construction of 
the Kraus family $\{W_j^{(k)}(\omega)\}_{j,k,\omega}$ shows that its linear span coincides with the linear span 
of the family ${\cal V}=\{V_j^{(k)}(\omega)\}_{j,k,\omega}$. By Eq. (\ref{KrausFam}), the family ${\cal V}$ is self-adjoint, 
and von Neumann's bicommutant theorem implies that the smallest subalgebra of ${\cal O}$ containing ${\cal V}$ is 
the bicommutant ${\cal V}^{\prime\prime}$. As shown by Spohn (see Theorem~3 in \cite{Sp2}), the condition 
${\cal V}^{\prime\prime}=\cO$ is equivalent to (\ref{spohn-cond}).
Hence, assuming strict positivity of the matrices $h_j(\omega)$ for all $j$ and $\omega$, Theorems~\ref{prelim-thm1} 
and~\ref{prelim-thm3} imply that the Spohn condition (\ref{spohn-cond}) is also sufficient  for Hypothesis (ER) to hold.

Note that 
\[\sigma(\rho)=\sum_{j=1}^M \sigma_j(\rho),\]
where $\sigma_j(\rho)$ is the entropy production of the  system ${\cal S}$ interacting only with the reservoir ${\cal R}_j$ 
via the Lindbladian ${\cal L}_j$.  If the matrix $h_j(\omega)$ is strictly positive and 
\begin{equation}
\label{spohn-cond-2}
\{Q_j^{(k)}\}_k^\prime\cap \{H_\cS\}^\prime=\cc \one,
\end{equation}
then, as discussed above, the QDS $\{\e^{t \cL_j}\}_{t\geq 0}$ is positivity improving. Moreover,
$\cL_j^\ast(\rho)=0$ iff $\rho=\e^{-\beta_j H_\cS}/\tr (\e^{-\beta_j H_\cS})$.\footnote{The same 
conditions ensure that the terms $\rho_{\beta_0}(D_j(H_\cS, H_\cS))$ in Theorem~\ref{thm-LRT} (1) 
are strictly positive, providing of course that $H_\cS\not\in \cc \one$.} Hence, we arrive at the 
following elegant condition (see \cite{LS1, AS}) which ensures that $\sigma(\rho)>0$ for all states $\rho$: there exists a pair 
$j_1, j_2$ such that $\beta_{j_1}\not=\beta_{j_2}$, the relation (\ref{spohn-cond-2}) holds for $j=j_1, j_2$, and 
 the matrix 
$h_j(\omega)$ is strictly positive for all $\omega$ and $j=j_1, j_2$.

In conclusion, under very general and natural conditions the class of weak coupling limit QDS introduced in \cite{LS1} satisfies Hypotheses 
(ER), (DB), (TR), (\KMSB), and has strictly positive entropy production.\footnote{At the current level of 
generality, the verification of Hypothesis (RE)  requires supplementing Davies' conditions with additional regularity assumptions which we shall not discuss for reasons of space. 
In practice, {\sl i.e.} in the context of concrete models, the verification of (RE) is typically an easy exercise.
}
Starting with  the seminal paper \cite{LS1}, such semigroups 
have been one of the basic paradigms of non-equilibrium quantum statistical mechanics.

%%%%%%%%%%%%%%%%%%%%%
\section{Full counting statistics}
\label{sect-FCS}
%%%%%%%%%%%%%%%%%%%
In this section, we elucidate the physical meaning of the measure $P_\rho^t$ introduced in 
Theorem~\ref{thm-1} in cases where the Lindbladian $\cL$ describes a weakly coupled open quantum 
system as discussed in the preceding section. We shall keep our presentation at a formal level; the 
interested reader should consult Section 5 of \cite{JOPP} for a more detailed discussion as well as
\cite{DdRM, dRK1, dRK2, JPPW} for a mathematically rigorous treatment of some specific models.

We start with the open system described in Section~\ref{sect-WCL}, but we assume now that the 
reservoirs $\cR_j$ are confined to finite boxes. More precisely, working in the Schr\"odinger 
representation, we assume that the reservoir Hamiltonians $H_j$ have purely discrete spectrum 
and that the operators $\e^{-\beta H_j}$ are trace class for all $\beta>0$. The initial state of the 
combined system is $\rho=\rho_\cS\otimes\rho_\cR$, where
$$
\rho_\cR=\rho_{\cR_1}\otimes\cdots\otimes\rho_{\cR_M},\qquad
\rho_{\cR_j}=\frac{\e^{-\beta_jH_j}}{\tr(\e^{-\beta_jH_j})},
$$
and $\rho_\cS$ is the initial state of the small system $\cS$.

The full counting statistics of the entropy fluxes across the system $\cS$ is defined as follows. 
Set $S= (S_1, \cdots, S_M)$ with $S_j=\beta_j H_j$. The observables $S_j$ commute and hence can
be simultaneously measured. Let $\Pi_s$  denote the joint spectral projection of $S$ associated to the eigenvalue $s\in \sp(S)$. Two 
successive measurements of $S$ at time $t_0$ and at time $t_0+t$ are described by the positive 
map valued measure (PMVM) (see, e.g., \cite{Da4}) which, to any subset $A\in\sp(S)\times\sp(S)$, 
associate the CP map
$$
{\cal E}_A(X)=\sum_{(s,s')\in A}\Pi_{s'}\e^{-\i tH_\lambda}\Pi_s X\Pi_s\e^{\i tH_\lambda}\Pi_{s'}.
$$
Indeed, if $\rho_{t_0}$ denotes the state of the system at time $t_0$, one easily checks that, 
according to the usual rules of projective measurements,
$$
\tr\,{\cal E}_{\{(s,s')\}}(\rho_{t_0}),
$$
is the joint probability for the first measurement to yield the result $s$ and for the second one to
yield the result $s'$. Hence, the probability distribution of $\varsigma=(s^\prime-s)/t$, the mean rate
of entropy transport from the system $\cS$ to the $M$ reservoirs over the time interval $[0, t]$, 
is given in terms of the initial state $\rho_\cS$ by the formula
\[
{\mathbb P}^t_{\rho_\cS}(\varsigma)=\tr\,{\cal E}_{\{s^\prime-s=t\varsigma\}}(\rho_\cS\otimes\rho_\cR).
\]
The atomic probability measure ${\mathbb P}^t_{\rho_\cS}$ on $\rr^M$ is the {\em full counting 
statistics} of the energy/entropy flow. An elementary calculation shows that the Laplace transform
of this measure is given by
\[
\ell^t_{\rho_\cS}(\alpha)
=\int_{\rr^M}\e^{-t\alpha \cdot \varsigma}\d{\mathbb P}^t_{\rho_\cS}(\varsigma)
=\tr\left((\rho_\cS\otimes\rho_\cR)\rho_\cR^{-\alpha}\e^{\i t H_\lambda}\rho_\cR^{\alpha}\e^{-\i t H_\lambda}\right),
\]
where, for $\alpha=(\alpha_1,\ldots,\alpha_M)\in\rr^M$, we have set
\[
\rho_\cR^\alpha
=\one\otimes\rho_{\cR_1}^{\alpha_1}\otimes\cdots\otimes\rho_{\cR_M}^{\alpha_M}.
\]
Assuming that  the operators
$$
\tau_j^{\i s\beta_j/2}(R_j^{(k)})=\e^{-s\beta_jH_j/2}R_j^{(k)}\e^{s\beta_jH_j/2},
$$
are entire analytic functions of $s$, we can define the deformed Hamiltonian
$$
H_{\lambda,\alpha}=\rho_\cR^{\alpha/2}H_{\lambda}\rho_\cR^{-\alpha/2}
=H_\cS+\sum_{j=1}^M\left(H_j
+\lambda^2\sum_{k=1}^{n_j}Q_j^{(k)}\otimes\tau_j^{\i\alpha_j\beta_j/2}(R_j^{(k)})\right),
$$
and write
\beq\label{der-1}
\ell^t_{\rho_\cS}(\alpha)
=\tr\left((\rho_\cS\otimes\rho_\cR)\,\e^{\i t H_{\lambda,\alpha}^\ast}\one\e^{-\i t H_{\lambda,\alpha}} \right).
\eeq
At this point, one can pass to the GNS representation of the reservoirs and perform a thermodynamic 
limit, letting the size of the confining boxes become infinite. If the deformed  operators 
$\tau_j^{\i \alpha_j\beta_j/2}(R_j^{(k)})$ remain well defined elements of the $W^\ast$-algebras 
$\cO_j$ in this limit, then we can define the effective deformed dynamics of the open system
with infinitely extended reservoirs
\beq\label{der-2}
\langle\psi|\cT^t_{\lambda,\alpha}(X)\psi\rangle
=\langle\psi\otimes\xi|\e^{\i t L_{\lambda,\alpha}^\ast}
(X\otimes \one)\e^{-\i t L_{\lambda,\alpha}}\psi\otimes\xi\rangle,
\eeq
with the deformed semi-standard Liouvillean
$$
L_{\lambda,\alpha}=H_\cS+\sum_{j=1}^M\left(L_j
+\sum_{k=1}^{n_j}Q_j^{(k)}\otimes\tau_j^{\i\alpha_j\beta_j/2}(R_j^{(k)})\right).
$$

Assuming that the thermodynamic limit
$$
{\rm TD-}\lim
\tr\left((|\psi\rangle\langle\psi|\otimes\rho_\cR)\,
\e^{\i t H_{\lambda,\alpha}^\ast}(X\otimes \one)\e^{-\i t H_{\lambda,\alpha}} \right)
=\langle\psi|\cT^t_{\lambda,\alpha}(X)\psi\rangle,
$$
exists for any $\psi\in\cH_\cS$, $X\in\cB(\cH_\cS)$, and $\alpha\in\rr^M$, we conclude
that the Laplace transform $\ell^t_{\rho_\cS}(\alpha)$ of the full counting statistics 
$\mathbb{P}^t_{\rho_\cS}$ has a well defined thermodynamic limit
$$
{\rm TD-}\lim\ell^t_{\rho_\cS}(\alpha)=\tr\left(\rho_\cS\cT^t_{\lambda,\alpha}(\one)\right),
$$
for all $\alpha\in\rr^M$. Then one can show that, as the size of the reservoir increases, the full 
counting statistics $\mathbb{P}^t_{\rho_\cS}$ converges weakly to a Borel probability measure 
which we again denote by $\mathbb{P}^t_{\rho_\cS}$  which satisfies
$$
\int_{\rr^M}\e^{-t\alpha\cdot\varsigma}\d\mathbb{P}^t_{\rho_\cS}(\varsigma)
=\tr\left(\rho_\cS\cT^t_{\lambda,\alpha}(\one)\right),
$$
(see Proposition~4.1 in \cite{JOPS}). We call the limiting measure $\mathbb{P}^t_{\rho_\cS}$ the
{\sl full counting statistics} of the open system $\cS$ coupled to the infinitely extended
reservoirs $\cR_1,\ldots,\cR_M$. Note that since infinitely extended reservoirs have an infinite
energy, it is not possible to implement directly the successive measurement procedure we have 
described to this model, and that one is forced to invoke the thermodynamic limit to construct its
full counting statistics.

Applying the Davies procedure to extract the weak coupling limit of the deformed effective dynamics leads to 
$$
\lim_{\lambda\to0}\sup_{\lambda^2t\in I}\|\cT^t_{\lambda,\alpha}-\e^{t(\cL_\cS+\lambda^2\cK_{(\alpha)})}\|=0,
$$
where  $\cK_{(\alpha)}$ is a deformed generator commuting with $\cL_\cS$. An explicit calculation shows 
that  the only difference between $\cK_{(\alpha)}$ and  the undeformed Lindbladian $\cL=\cK_0$ is that  the functions 
$h_j^{(kl)}$ are replaced with (recall that $L_j\xi_j=0$),
\begin{align*}
h_{j,\alpha_j}^{(kl)}(\omega)
&=2\pi\langle\tau_j^{\i\alpha_j\beta_j/2}(R^{(k)}_j)\xi_j|
\delta(L_j-\omega)\tau_j^{\i\alpha_j\beta_j/2}(R^{(l)}_j)\xi_j\rangle\\
&=2\pi\langle\e^{-\alpha_j\beta_jL_j/2}R^{(k)}_{j}\xi_j|
\delta(L_j-\omega)\e^{-\alpha_j\beta_jL_j/2}R^{(l)}_{j}\xi_j\rangle\\
&=\e^{-\alpha_j\beta_j\omega}h_{j}^{(kl)}(\omega).
\end{align*}
Using Eq.~\eqref{KrausFam-1}, one finally concludes that, with the $\rho_j$ defined in Eq.~\eqref{rho-j-eq}, 
$$
\cK_{(\alpha)}(X)=\sum_{j=1}^M\cL_j(X\rho_j^{-\alpha_j})\rho_j^{\alpha_j},
$$
and so $\cK_{(\alpha)}$  coincides with the deformed Lindbladian $\cL_{(\alpha)}$. We conclude that   if $[H_\cS, \rho_\cS]=0$, then 
the  measure
$P_\rho^t$ introduced in Theorem~\ref{thm-1} is related to the full counting statistics
$\mathbb{P}^t_{\rho_\cS}$ through the scaling limit
\beq
\lim_{\lambda\to0}\int f(\lambda^{-2}\varsigma)\d\mathbb{P}^{t/\lambda^2}_{\rho_\cS}(\varsigma)=
\int f(\varsigma)\d P^t_{\rho_\cS}(\varsigma).
\label{FCS-limit}
\eeq
We note that in the weak coupling regime, the energy/entropy fluxes are of order
$\lambda^2$ so the scaling $\lambda^{-2}\varsigma$ which appears on the left hand side of
the last identity is natural. The measure $P^t_{\rho_\cS}$ thus describes the rescaled
mean energy/entropy fluxes at the Van Hove time scale $t/\lambda^2$. To the best of our knowledge, 
this observation is  due to de~Roeck \cite{dR1}.

For some specific models it is possible to show that 
\[
\lim_{\lambda\to0}\sup_{t>0}\|\cT^t_{\lambda,\alpha}-\e^{t(\cL_\cS+\lambda^2\cK_\alpha)}\|=0,
\]
(see \cite{dR2, dRK1, dRK2, JPPW}), and  in such cases one can  relate the large deviation 
principle of Corollary~\ref{prop-LDP} to the large deviation principle of the full counting statistics 
${\mathbb P}_{\rho_\cS}^t$. 

The link between full counting statistics and deformations of  the semi-standard Liouvillean (relations (\ref{der-1}) and 
(\ref{der-2})) goes back to \cite{dR1}. The link between full counting statistics and deformations of the standard Liouvillean 
can be traced back to \cite{JP1, MT, DJ}, was fully elaborated in \cite{JOPP}, and plays the key role in the work \cite{JPPW}. The second  link 
relates full counting statistics to modular theory of operator algebras and deformed Lindbladians ${\cal L}_{(\alpha)}$ to Fermi Golden Rule 
for spectral resonances of the deformed standard Liouvilleans. This  point  is discussed in detail in 
\cite{JPPW} and we refer the reader to this work for additional information.

%%%%%%%%%%%%%%%%%%%%%%%%%%%%%%%%%
\section{Unraveling of the deformed semigroup $\e^{t\cL_{(\alpha)}}$}
\label{sec-unravel}

In this section we follow \cite{dRM} and present an alternative and more general interpretation of 
the measure $P^t_{\rho}$ based on the standard unraveling technique. As a byproduct of this 
construction, we shall get a 
proof of Parts (1) and (2) of Theorem~\ref{thm-1}. We shall assume that Hypothesis (DB) holds
throughout the section and use the elementary properties of Lindbladians summarized in 
Theorem~\ref{thm-DB}.

Let $\cL_j(X)=\i[T_j,X]-\frac12\{\Phi_j(\one),X\}+\Phi_j(X)$ denote a Lindblad decomposition of $\cL_j$
and set
$$
\cK(X)=-K^\ast X-XK,\qquad
K=\sum_{j=1}^MK_j,\qquad
K_j=\frac12\Phi_j(\one)+\i T_j.
$$
By Theorem~\ref{thm-DB}, $K_j$ commutes with $\rho_j$ and $\Phi_j$ admits a decomposition
$$
\Phi_j=\sum_{\omega\in\Omega_j}\Phi_{j,\omega},
$$
where $\Omega_j=\{\mu-\nu\,|\,\mu,\nu\in\sp(\log\rho_j)\}$ and $\Phi_{j,\omega}\in\CP(\cO)$
satisfies $\Phi_{j,\omega}(X\rho_j^{-\alpha_j})\rho_j^{\alpha_j}=\e^{-\alpha_j\omega}\Phi_{j,\omega}(X)$.
It follows that
\beq
\cL_{(\alpha)}
%=\sum_{j=1}^M\cR_{\rho_j}^{\alpha_j/2}\circ\cL_j\circ\cR_{\rho_j}^{-\alpha_j/2}
=\cK+\sum_{j=1}^M\sum_{\omega\in\Omega_j}\e^{-\alpha_j\omega}\Phi_{j,\omega},
\label{Lalpha-dec}
\eeq
is of the Lindblad form (\ref{prelim-4}) for $\alpha\in\rr^M$, which proves Part (1) of 
Theorem~\ref{thm-1}. Using the Dyson expansion of the cocycle 
$\Gamma^t_{\alpha}=\e^{-t\cK}\circ\e^{t\cL_{(\alpha)}}$, we obtain the representation
\begin{align}
\langle\rho&|\e^{t\cL_{(\alpha)}}(\one)\rangle=\langle\e^{t\cK^\ast}(\rho)|\Gamma^t_\alpha(\one)\rangle
=\langle\rho_t|\one\rangle\label{unraveq}\\
&+\sum_{N\ge1}
\sum_{\substack{(j_1,\ldots,j_N)\in\{1,\ldots,M\}^N\!\!
\\(\omega_1,\ldots,\omega_N)\in\Omega_{j_1}\times\cdots\times\Omega_{j_N}}}\!\!\!\!
\e^{-\sum_{k=1}^N\alpha_{j_k}\omega_k}\!\!\!\!\!\!
\int\limits_{0\le s_1\le\cdots\le s_N\le t}\!\!\!\!\!\!\!\!
\langle\rho_t|\Phi_{j_N,\omega_N,s_N}\circ\cdots\circ\Phi_{j_1,\omega_1,s_1}(\one)\rangle
\d s_1\cdots\d s_N,\nonumber
\end{align}
where $\rho_t=\e^{t\cK^\ast}(\rho)$ and 
$\Phi_{j,\omega,s}=\e^{-s\cK}\circ\Phi_{j,\omega}\circ\e^{s\cK}$.

Unraveling consists of rewriting this expression in terms of a probability measure $\mu^t_\rho$ on a
set $\Xi^t$ of quantum trajectories defined as follows. For $N\ge1$, let
$$
\Xi_N^t=
\{\xi=[\xi_1,\ldots,\xi_N]\,|\,\xi_k=(j_k,\omega_k,s_k),j_k\in\{1,\ldots,M\},
\omega_k\in\Omega_{j_k},0\le s_1\le\cdots\le s_N\le t\},
$$
and set $\Xi^t_0=\{\emptyset\}$. On the disjoint union
$$
\Xi^t=\bigsqcup_{N\ge0}\Xi_N^t,
$$
one defines the positive measure $\mu^t_\rho$ by
\begin{align*}
\int_{\Xi^t}&f(\xi)\,\d\mu^t_\rho(\xi)
=f(\{\emptyset\})\langle\rho_t|\one\rangle
+\sum_{N\ge1}\sum_{\substack{(j_1,\ldots,j_N)\in\{1,\ldots,M\}^N
\\(\omega_1,\ldots,\omega_N)\in\Omega_{j_1}\times\cdots\times\Omega_{j_N}}}\\
&\int\limits_{0\le s_1\le\cdots\le s_N\le t}f([(j_1,\omega_1,s_1),\ldots,(j_N,\omega_N,s_N)])
\langle\rho_t|
\Phi_{j_N,\omega_N,s_N}\circ\cdots\circ\Phi_{j_1,\omega_1,s_1}(\one)\rangle\,\d s_1\cdots\d s_N.
\end{align*}
Comparison with Eq.~\eqref{unraveq} shows that
$$
\int_{\Xi^t}\,\d\mu^t_\rho(\xi)=\langle\rho_t|\Gamma^t_0(\one)\rangle=\langle\rho|\e^{t\cL}(\one)\rangle=1,
$$
and hence $\mu^t_\rho$ is a probability measure. An element $\xi\in\Xi^t$ is a quantum trajectory which
represent the history of the system during  the time interval $[0,t]$.   Observe that the system can exchange entropy with the reservoir
$\cR_j$ only in quanta of the form $\omega\in\sp(S_j)-\sp(S_j)=\Omega_j$ where $S_j=-\log\rho_j$. An element 
$\xi=[\xi_1,\ldots,\xi_N]$ of $\Xi^t$ is a chronologically ordered list of elementary events
$\xi_k=(j_k,\omega_k,s_k)$ which we interpret in the following way: at time $s_k$ the system
has exchanged an entropy quantum $\omega_k$ with reservoir $\cR_{j_k}$. According to this
interpretation, the random variable
$$
\varsigma_j(\xi)=\frac1t\sum_{k:j_k=j}\omega_k,
$$
represents the mean rate of entropy exchange of the system with reservoir $\cR_j$ during the
time interval $[0,t]$. It follows that  one can rewrite the expansion~\eqref{unraveq} as
$$
\langle\rho|\e^{t\cL_{(\alpha)}}(\one)\rangle
=\int_{\Xi^t}\e^{-t\sum_{j=1}^M\alpha_j\varsigma_j(\xi)}\,\d\mu^t_\rho(\xi).
$$
This proves Part (2) of Theorem~\ref{thm-1} and identifies the measure $P_\rho^t$ as the
law of the random variable $\varsigma(\xi)=(\varsigma_1(\xi),\ldots,\varsigma_M(\xi))$ induced by the
measure $\mu^t_\rho$.

%%%%%%%%%%%%%%%%%%%%%%%%%%%%%%%%%%%%%%%%%%%%%%%
%%%%%%%%%%%%%%%%%%%%%%%%%%%%%%%%%%%%%%%%%%%%%%%

\section{Proofs}
\label{sect-proofs}
%%%%%%%%%%%%%%%%%%%%%%%%%%%%%%%%%%%%%%%%%%%%%%%
%%%%%%%%%%%%%%%%%%%%%%%%%%%%%%%%%%%%%%%%%%%%%%%
\subsection{Detailed balance}

To a faithful state $\rho$, we associate two groups of transformations of $\cO$, the modular group 
$\Delta_\rho^z(X)=\rho^zX\rho^{-z}$, and the group  $\cR_\rho^z(X)=\rho^z X\rho^z$, $z\in\cc$. 
$\Delta_\rho^1=\Delta_\rho$ is the modular operator of the state $\rho$. 
Note that $\Delta_\rho^{\i\alpha}\in\CP_1(\cO)$ and $\cR_\rho^\alpha\in\CP(\cO)$ for $\alpha\in\rr$.

\bet\label{thm-DB} 
Let $\rho$ be a faithful state on $\cO$ and $\cL=\i[T,\,\cdot\,]-\frac12\{\Phi(\one),\,\cdot\,\}+\Phi$  
a Lindbladian generating a QDS. Suppose  that $\cL^\ast(\rho)=0$ and 
$\Phi^\rho=\Phi$. Then:
\begin{enumerate}[{\rm (1)}]
\item The Hermitian and anti-Hermitian parts of $\cL$ w.r.t.\;the inner product induced by $\rho$
are given by
\[
\begin{split}
\cL^{(d)}(X)&=\frac12(\cL+\cL^\rho)(X)=-\frac12\{\Phi(\one),X\}+\Phi(X),\\[3mm]
\cL^{(h)}(X)&=\frac12(\cL-\cL^\rho)(X)=\i[T,X].
\end{split}
\]
They are also called dissipative and Hamiltonian parts of $\cL$.
\item $\cL$, $\cL^{(h)}$, $\cL^{(d)}$ and $\Phi$ commute with the modular operator $\Delta_\rho$.
In particular, $T$ and $\Phi(\one)$ commute with $\rho$ and $\cL^{(d)\ast}(\rho)=\cL^{(h)\ast}(\rho)=0$.
\item The CP map $\Phi$ admits a decomposition
$$
\Phi=\sum_{\omega\in\sp(\log\Delta_\rho)}\Phi_\omega,
$$
where $\Phi_\omega\in\CP(\cO)$ satisfies 
$\Phi_\omega(X\rho^{-\alpha})\rho^\alpha=\e^{-\alpha\omega}\Phi_\omega(X)$,
$\Phi_\omega^\rho=\Phi_{-\omega}$ and $\Phi_\omega^\ast=\e^\omega\Phi_{-\omega}$.
\item For $\alpha\in\cc$ define $\cL_\alpha=\cR_\rho^{\alpha/2}\circ\cL\circ\cR_\rho^{-\alpha/2}$.
Then
\beq\label{utah-airport-1}
\cL_\alpha(X)=\cL(X\rho^{-\alpha})\rho^\alpha
=\i[T,X]-\frac12\{\Phi(\one),X\}
+\sum_{\omega\in\sp(\log\Delta_\rho)}\e^{-\alpha\omega}\Phi_\omega(X),
\eeq
holds for all $X\in\cO$. Moreover, $\{\e^{t\cL_\alpha}\}_{t\geq 0}$ is a $\CP(\cO)$ semigroup for  $\alpha\in\rr$.
\item If the pair  $(\rho, \cL)$ is time-reversible with time-reversal $\Theta$, then for all $\alpha \in \rr$
\[\Theta\circ\cL_\alpha^\ast=\cL_{1-\alpha}\circ\Theta.\]
\end{enumerate}
\eet
{\bf Remark.}  The proofs of Parts (1)-(3) can be found in \cite{Al, KFGV}. For the readers convenience we provide a complete 
proof below. 

\demo We start with the simple remarks that $\cL^\ast(X)=-\i[T,X]-\frac12\{\Phi(\one),X\} +\Phi^\ast(X)$ 
and that $\cM^\rho(X)=\cM^\ast(X\rho)\rho^{-1}$ for any linear map $\cM$ on $\cO$. We recall
that $\cM$ is a $\ast$-map if $\cM(X^\ast)=\cM(X)^\ast$ for all $X\in\cO$. The maps 
$\cL$ and $\cL^\ast$, as generators of positive semigroups, and $\Phi$ as a positive map,
are all $\ast$-maps.

The fact that $\Phi$ is 
$\rho$-self-adjoint implies $\Phi^\ast(X\rho)=\Phi(X)\rho$ for all $X\in\cO$ and in particular that 
$\Phi^\ast(\rho)=\Phi(\one)\rho$. Thus, since $T$, $\Phi(\one)$ and $\rho$ are self-adjoint, it follows from
$$
0=\cL^\ast(\rho)=-\i[T,\rho]+\Phi^\ast(\rho)-\frac12\{\Phi(\one),\rho\}=\frac12[\Phi(\one)-2\i T,\rho],
$$
that $\rho$ commutes with $T$ and $\Phi(\one)$. A simple calculation  yields
\beq\label{utah-airport}
\cL^\rho(X)=-\i[T,X]+\Phi(X)-\frac12\{\Phi(\one),X\},
\eeq
and Part (1) follows. 

The formula  (\ref{utah-airport}) implies that $\cL^\rho$ is a $\ast$-map. Thus, one can write
\begin{equation}\label{utah}
\begin{split}
\cL^\rho(\Delta_\rho(X))&=\cL^\ast(\rho X)\rho^{-1}=\cL^\ast((X^\ast\rho)^\ast)\rho^{-1}
=\cL^\ast(X^\ast\rho)^\ast\rho^{-1}\\[3mm]
&=(\cL^\rho(X^\ast)\rho)^\ast\rho^{-1}=\Delta_\rho(\cL^\rho(X)).
\end{split}
\end{equation}
It follows that $[\cL^\rho,\Delta_\rho]=0$ and, since $\Delta_\rho$ is $\rho$-self-adjoint,  that $[\cL,\Delta_\rho]=0$.
Clearly,  $[T,\rho]=0$ implies that $[\cL^{(h)},\Delta_\rho]=0$ and 
$\cL^{(h)\ast}(\rho)=0$. Thus, one also has $[\cL^{(d)},\Delta_\rho]=0$ and 
\[\cL^{(d)\ast}(\rho)=\cL^\ast(\rho)-\cL^{(h)\ast}(\rho)=0.\]
Finally,
$[\Phi(\one),\rho]=0$ implies $[\Phi,\Delta_\rho]=0$, which concludes the proof of 
Part (2).

Denote by $\log\rho=\sum_\lambda\lambda P_\lambda$ the spectral representation of
$\log\rho$. The operator $\log\Delta_\rho=[\log\rho,\,\cdot\,]$ is self-adjoint on $\cO$, with spectrum
$\sp(\log\Delta_\rho)=\sp(\log\rho)-\sp(\log\rho)$. Its spectral representation is given by
$$
\log\Delta_\rho=\sum_{\omega\in\sp(\log\Delta_\rho)}\omega\cP_\omega,
\qquad
\cP_\omega(X)=\sum_{\lambda-\mu=\omega}P_\lambda XP_\mu.
$$
Since $\Phi$ commutes with $\Delta_\rho$, it commutes with each of the spectral
projection $\cP_\omega$, and in particular one has
$\cP_\omega\circ\Phi\circ\cP_\omega=\cP_\omega\circ\Phi$.
Thus, setting
$$
\Phi_\omega(X)=\sum_{\substack{\lambda-\mu=\omega\\\lambda'-\mu'=\omega}}
P_{\mu}\Phi(P_\lambda XP_{\lambda'})P_{\mu'},
$$
we have
\begin{align*}
\sum_{\omega\in\sp(\log\Delta_\rho)}\Phi_\omega(X)
&=\sum_{\lambda-\mu=\lambda'-\mu'}P_{\mu}\Phi(P_\lambda XP_{\lambda'})P_{\mu'}
=\sum_{\lambda-\lambda'=\mu-\mu'}P_{\mu}\Phi(P_\lambda XP_{\lambda'})P_{\mu'}\\
&=\sum_{\omega\in\sp(\log\Delta_\rho)}\cP_\omega(\Phi(\cP_\omega(X)))
=\sum_{\omega\in\sp(\log\Delta_\rho)}\cP_\omega(\Phi(X))=\Phi(X).
\end{align*}
Moreover, since $\Phi$ is completely positive, it follows from the identity
$$
\sum_{i,j}B_i^\ast\Phi_\omega(A_i^\ast A_j)B_j=\sum_{(i,\lambda),(j,\mu)}
B_{i,\lambda}^\ast\Phi(A_{i,\lambda}^\ast A_{j,\mu})B_{j,\mu},
$$
where $A_{i,\lambda}=A_iP_\lambda$ and $B_{i,\lambda}=P_{\lambda-\omega}B_i$, that
$\Phi_\omega$ is completely positive. Next, note that the identity
$$
P_{\mu}\Phi(P_\lambda X\rho^{-\alpha}P_{\lambda'})P_{\mu'}\rho^{\alpha}
=P_{\mu}\Phi(P_\lambda XP_{\lambda'})P_{\mu'}\e^{-\alpha(\lambda'-\mu')},
$$
implies 
\beq
\Phi_\omega(X\rho^{-\alpha})\rho^{\alpha}=\e^{-\alpha\omega}\Phi_\omega(X).
\label{PhiBeauty}
\eeq
The identity  $\Phi^\ast(X)=\Phi(X\rho^{-1})\rho$  and a simple calculation yield
\begin{align*}
\Phi_\omega^\ast(X)=\sum_{\substack{\lambda-\mu=\omega\\\lambda'-\mu'=\omega}}
P_{\lambda}\Phi^\ast(P_\mu XP_{\mu'})P_{\lambda'}
&=\sum_{\substack{\lambda-\mu=\omega\\\lambda'-\mu'=\omega}}
P_{\lambda}\Phi(P_\mu XP_{\mu'}\rho^{-1})\rho P_{\lambda'}\\
&=\sum_{\substack{\lambda-\mu=\omega\\\lambda'-\mu'=\omega}}
P_{\lambda}\Phi(P_\mu XP_{\mu'})P_{\lambda'}\e^{\lambda'-\mu'}=\e^{\omega}\Phi_{-\omega}(X).
\end{align*}
The last identity combined with Eq.~\eqref{PhiBeauty} gives
$$
\Phi_\omega^\rho(X)=\Phi_\omega^\ast(X\rho)\rho^{-1}=\e^\omega\Phi_{-\omega}(X\rho)\rho^{-1}
=\Phi_{-\omega}(X),
$$
and Part (3) follows.

To prove Part (4), note that since $\cL$ commutes with $\Delta_\rho$, one has
\begin{align*}
\cL_\alpha(X)
&=\cR_\rho^{\alpha/2}\circ\cL\circ\cR_\rho^{-\alpha/2}(X)
=\cR_\rho^{\alpha/2}\circ\Delta_\rho^{-\alpha/2}\circ\cL
\circ\Delta_\rho^{\alpha/2}\circ\cR_\rho^{-\alpha/2}(X)\\[4pt]
&=\rho^{\alpha/2}\rho^{-\alpha/2}
\cL(\rho^{\alpha/2}\rho^{-\alpha/2}X\rho^{-\alpha/2}\rho^{-\alpha/2})
\rho^{\alpha/2}\rho^{\alpha/2}=\cL(X\rho^{-\alpha})\rho^\alpha.
\end{align*}
The formula  (\ref{utah-airport-1}) follows from the relation $\cL_\alpha(X)=\cL(X\rho^{-\alpha})\rho^\alpha$,
the fact that $\rho$ commutes with $T$ and $\Phi(\one)$ and Eq.~\eqref{PhiBeauty}.
Since $\e^{t\cL_\alpha}=\cR_\rho^{\alpha/2}\circ\e^{t\cL}\circ\cR_\rho^{-\alpha/2}$, 
$\{\e^{t\cL_\alpha}\}_{t\geq 0}$ is a $\CP(\cO)$ semigroup for all $\alpha \in \rr$, and Part (4) follows.

It remains to prove Part (5).  Define $\cL_\alpha^{(d)}(X)=\cL^{(d)}(X\rho^{-\alpha})\rho^\alpha$. A simple calculation gives
\[\cL_\alpha^{(d)\ast}(X)=\cL^{(d)\ast}(X\rho^{\alpha})\rho^{-\alpha},\]
and Part (1) implies
\[\cL^{(d)\ast}(X)=\cL^{(d)\rho}(X\rho^{-1})\rho=\cL^{(d)}(X\rho^{-1})\rho.\] Hence,
$$
\cL_\alpha^{(d)\ast}=\cL_{1-\alpha}^{(d)}.
$$

Since $\Theta$ is involutive, the relation $\cL^\rho\circ\Theta=\Theta\circ\cL$ implies
$\cL\circ\Theta=\Theta\circ\cL^\rho$. It follows from Part (1) that 
$\cL^{(h)}\circ\Theta=-\Theta\circ\cL^{(h)}$ and  $\cL^{(d)}\circ\Theta=\Theta\circ\cL^{(d)}$. 
Moreover, $\Theta(\rho^\alpha)=\rho^\alpha$
implies $\cL_\alpha^{(d)}\circ\Theta=\Theta\circ\cL_\alpha^{(d)}$. Thus, one has
$$
\cL_\alpha^\ast=\cL^{(h)\ast}+\cL_\alpha^{(d)\ast}=-\cL^{(h)}+\cL_{1-\alpha}^{(d)},
$$
and 
$$
\Theta\circ\cL_\alpha^\ast
=(\cL^{(h)}+\cL_{1-\alpha}^{(d)})\circ\Theta=\cL_{1-\alpha}\circ\Theta.
$$
\qed

We finish this section with:
\bet\label{utah-knee}
Let $\rho$ be a faithful state and $\cL$ a Lindbladian on $\cO$ generating a QDS. Suppose that 
$\cL^\ast(\rho)=0$. Then the following statements are equivalent: 
\begin{enumerate}[{\rm (1)}]
\item
There exist  a self-adjoint $T\in \cO$ such that the Hermitian part of $\cL$ w.r.t.\;the inner product induced 
by $\rho$ has the form
\[\cL^{(h)}(X)=\frac{1}{2}(\cL- \cL^\rho)(X)=\i [T, X].\]
\item There exists a Lindblad decomposition $\cL=\i[T,\,\cdot\,]-\frac12\{\Phi(\one),\,\cdot\,\}+\Phi$ such that 
$\Phi^\rho=\Phi$.
\end{enumerate}
\eet
{\bf Remark.} This theorem establishes that Definition \ref{prelim-def} (a) is equivalent to the definition of detailed 
balance given in \cite{KFGV} (see also Section IV in \cite{LS1}). Although we shall not make use of this result in the sequel, 
we include the proof for reader's convenience. 

\demo The implication (2) $\Rightarrow$ (1) follows from Part (1) of Theorem~\ref{thm-DB}.  To prove the implication 
(1) $\Rightarrow$ (2), note first that (1) implies that 
\[
\cL^{(d)}=\frac{1}{2}(\cL + \cL^{\rho}),
\]
is a Lindbladian generating a QDS. Since ${\cal L}^{(d)}$ is $\rho$-self-adjoint, arguing as in (\ref{utah}) one deduces that 
$[\Delta_\rho, \cL^{(d)}]=0$. Let now $\cL=\i[S,\,\cdot\,]-\frac12\{\Psi(\one),\,\cdot\,\}+\Psi$ be a Lindblad decomposition. Since 
\[
{\cal L}^{(d)}=\lim_{T\rightarrow \infty}\frac{1}{T}\int_0^T \Delta_{\rho}^{\i t}\circ \cL^{(d)}\circ \Delta_{\rho}^{-\i t}\d t,
\]
setting 
\[ M=\lim_{T\rightarrow \infty} \frac{1}{T}\int_0^T \rho^{\i t}S \rho^{-\i t}\d t, \qquad 
\Xi=\lim_{T\rightarrow \infty}\frac{1}{T}\int_0^T \Delta_{\rho}^{\i t}\circ\Psi \circ \Delta_{\rho}^{-\i t}\d t,
\]
we deduce that 
\[
\cL^{(d)}= \i[M,\,\cdot\,]-\frac12\{\Xi(\one),\,\cdot\,\}+\Xi,
\]
is also a Lindblad decomposition. Clearly, 
$[\Delta_\rho, \Xi]=0$, $[M, \rho]=0$, $[\Xi(\one), \rho]=0$. Hence, 
\begin{align*}
\cL^{(d)\rho}&= -\i[M,\,\cdot\,]-\frac12\{\Xi(\one),\,\cdot\,\}+\Xi^\rho,\\
\Xi^{\rho}(X)&=\rho^{-1/2}\Xi^\ast (\rho^{1/2}X\rho^{1/2})\rho^{-1/2},
\end{align*}
and we derive that $\Xi^\rho(\one)=\Xi(\one) + \cL^{(d)\rho}(\one)=\Xi(\one) + \cL^{(d)}(\one)=\Xi(\one)$. Setting 
\[
\Phi=\frac{1}{2}(\Xi + \Xi^\rho),
\]
we get 
\[
\cL^{(d)}=\frac{1}{2}(\cL^{(d)} + \cL^{(d)\rho})=-\frac{1}{2}\{\Phi(\one),\,\cdot\,\} +\Phi,
\]
where $\Phi$ is CP and  $\Phi^\rho=\Phi$. Hence, $\cL=\i[T,\,\cdot\,]-\frac12\{\Phi(\one),\,\cdot\,\}+\Phi$ is a Lindblad 
decomposition of $\cL$ with $\Phi^\rho=\Phi$. \qed

%%%%%%%%%%%%%%%%%%%%%%%%%%%%%%%%
\subsection{Irreducibility and positivity improving}
\label{sect-AQDS}
%%%%%%%%%%%%%%%%%%%%%%
We start with the following observation of \cite{Sc}:
\bep\label{shrad}
A positive linear map $\Phi: 
\cO \rightarrow \cO$ is irreducible iff  $\e^{t \Phi}$ is positivity improving for some (and hence all) $t>0$. 
\eep
\demo  If $\Phi$ is irreducible, then it follows from Lemma~2.1 in \cite{EHK} that 
\[
({\rm Id} + \Phi)^{\dim \cH-1},
\]
is positivity improving, and so $\e^{t \Phi}$ is positivity improving for all $t>0$. To prove the converse, 
suppose that $\e^{t \Phi}$ is positivity improving and that $\Phi({P})\leq \lambda P$, where $\lambda >0$ and 
$P\not=0$ is a projection. Then $\Phi^n({P})\leq \lambda^n P$ for all $n$, and so $0<\e^{t \Phi}({P})\leq \e^{\lambda t}P$. The last 
relation implies that $P=\one$.\qed

{\bf Proof of Theorem~\ref{prelim-thm1}.} We follow \cite{Sc}. Let 
$\varphi,\psi\in\cH$ be non-zero vectors and $t>0$. Expanding 
$\e^{t\Phi^\ast}$ into a power series, we get 
\[
\langle\varphi|\e^{t\Phi^\ast}(|\psi\rangle\langle\psi|)\varphi\rangle=
|\langle \varphi|\psi\rangle|^2 +\sum_{n=1}^\infty\frac{t^n}{n!}\sum_{j_1, \cdots, j_n}|\langle \varphi|V_{j_1}\cdots V_{j_n}\psi\rangle|^2.
\]
Hence, $\langle\varphi|\e^{t\Phi^\ast}(|\psi\rangle\langle\psi|)\varphi\rangle=0$ iff $\varphi \perp {\cal A}\psi$, 
and we deduce that $\e^{t\Phi^\ast}$ is positivity improving iff ${\cal A}\psi=\cH$ for all non-zero vectors $\psi \in \cH$. 
Since $\e^{t\Phi^\ast}$ is positivity improving iff  $\e^{t\Phi}$ is, the result follows from Proposition~\ref{shrad}. \qed

{\bf Proof of Theorem~\ref{prelim-thm2}.} The proof of based on Perron-Frobenius theory of positive maps developed 
in \cite{EHK}.  Let $t>0$ be given.  The map $\e^{t \cL}$ is positive and its spectral radius is $\e^{t \ell}$. It  follows from  
Theorem~2.5 in \cite{EHK} that $\e^{t \ell}$ is an eigenvalue of $\e^{t \cL}$, and that there exists a non-zero 
$M\in \cO_+$ such that 
\[ \e^{t \cL}(M)=\e^{t \ell}M.
\]
Since the map $\e^{t \cL}$ is positivity improving, $M>0$. Define 
$$
\Psi(X)=M^{-1/2}\e^{t(\cL-\ell)}(M^{1/2}XM^{1/2})M^{-1/2}.
$$
The map $\Psi$ is unital, completely positive (hence Schwartz), and positivity improving (hence irreducible).
The same holds for $\Psi^n$ , $n\geq 1$, and it follows from Theorem~4.2 in \cite{EHK}  that $1$ is a simple eigenvalue of 
$\Psi$ and that $\Psi$ has no other eigenvalues on the unit circle $|z|=1$. Hence, $\cL$ has a simple eigenvalue 
at $\ell$ and  no other eigenvalues on the line $\Re z =\ell$. 

Denote by $\mu$ the eigenvector of $\cL^\ast$ associated to the 
eigenvalue $\ell$. Since $\e^{t\cL^\ast}$ is positivity improving by duality, one can chose
$\mu>0$ and normalize it by $\langle\mu|M\rangle=1$. Let $\delta>0$ be the distance
from $\sp(\cL)\setminus\{\ell\}$ to the line $\Re z=\ell$. Then, for any $\epsilon>0$,
\beq
\langle\rho|\e^{t\cL}(X)\rangle=\e^{t\ell}\left(\langle\rho|M\rangle\langle\mu|X\rangle
+O\left(\e^{-t(\delta-\epsilon)}\right)\right),
\label{ell-eq}
\eeq
holds for all states $\rho$ and all $X\in\cO$. Since $\langle\rho|M\rangle>0$ 
and $\langle\mu|X\rangle>0$ for non-zero $X\in\cO_+$, Eq.~\eqref{prelim-3} follows.

If $\cL(\one)=0$, then $\e^{t\cL}(\one)=\one$ and since $\|\e^{t\cL}\|=\|\e^{t\cL}(\one)\|=1$, it follows that
$\ell=0$ and $M=\one$. By duality, \eqref{ell-eq} yields
$$
\e^{t\cL^\ast}(\rho)=\mu+O(\e^{-t(\delta-\epsilon)}),
$$
and the semigroup $\{\e^{t \cL}\}_{t\geq 0}$ is relaxing exponentially fast to the faithful state $\rho_+=\mu$.
\qed

{\bf Proof of Theorem~\ref{prelim-thm3}.}  Note that $\cK_0\colon X\mapsto K^\ast X+XK$ generates a continuous
group of completely positive maps on $\cO$, namely $\e^{t\cK_0}(X)=\e^{tK^\ast}X\e^{tK}$.
Denoting $\Gamma^t=\e^{-t\cK_0}\circ\e^{t\cK}$, it is sufficient to show that 
$\langle\varphi|\Gamma^t(|\psi\rangle\langle\psi|)\varphi\rangle>0$ for any non-zero vectors
$\varphi,\psi\in\cH$ and all $t>0$. To prove this claim, let us assume that 
$\langle\varphi|\Gamma^{t_0}(|\psi\rangle\langle\psi|)\varphi\rangle=0$ for some $t_0>0$. 
The Dyson expansion for $\Gamma^{t_0}$ gives 
$$
\Gamma^{t_0}={\rm Id}+\sum_{n=1}^\infty\ \int_{\Delta_n}\Phi^n_s\,\d s,
$$
where
\[\Delta_n=\{s=(s_1,\ldots,s_n)\in\rr^n\,|\,0\le s_1\le\cdots\le s_n\le t_0\},\]
$\Phi_s=\e^{-s\cK_0}\circ\Phi\circ\e^{s\cK_0}$ for $s\in\rr$, and 
$\Phi^n_s=\Phi_{s_1}\circ\Phi_{s_2}\circ\cdots\circ\Phi_{s_n}$ for $s\in\Delta_n$. It follows that 
$$
\langle\varphi|\Gamma^{t_0}(|\psi\rangle\langle\psi|)\varphi\rangle=|\langle\varphi|\psi\rangle|^2
+\sum_{n=1}^\infty\ \int_{\Delta_n}
\langle\varphi|\Phi^n_s(|\psi\rangle\langle\psi|)\varphi\rangle\,\d s=0.
$$
Since the functions  $s\mapsto \langle\varphi|\Phi^n_s(|\psi\rangle\langle\psi|)\varphi\rangle$ are continuous and 
non-negative, we infer that  $\langle\varphi|\Phi^n_s(|\psi\rangle\langle\psi|)\varphi\rangle=0$ 
for all $n$ and all $s\in\Delta_n$, and in particular that 
$\langle\varphi|\Phi^n(|\psi\rangle\langle\psi|)\varphi\rangle=0$ for all $n$. Hence, 
$\langle\varphi|\e^{t\Phi}(|\psi\rangle\langle\psi|)\varphi\rangle=0$ for all $t\ge0$, and Proposition~\ref{shrad} implies that 
$\varphi=0$ or $\psi=0$. \qed 

For later reference, we mention the following simple fact:
\bep\label{prop-ergo}
Let $\Phi_j$, $j=1,\cdots,n$, be positive linear maps  such that $\sum_j \Phi_j$ is 
irreducible. If $\lambda_1,\ldots,\lambda_n$ are strictly positive then $\sum_j \lambda_j \Phi_j$ 
is irreducible.
\eep
\demo The result follows from the obvious inequality 
\[
\min_i \lambda_i \sum_j \Phi_j \leq \sum_j \lambda_j \Phi_j\leq \max_i \lambda_i \sum_j \Phi_j.
\]
\qed

\subsection{Proof of Theorem~\ref{thm-1}}
\label{subsect-proof-thm-1}

(1)-(2) were already proven in Section~\ref{sec-unravel}.

(3)-(4) By Eq.~\eqref{Lalpha-dec}, Proposition~\ref{prop-ergo}, and Theorem~\ref{prelim-thm3}, 
the CP semigroup  $\{\e^{t\cL_{(\alpha)}}\}_{t\geq 0}$ is positivity improving for all  $\alpha\in\rr^M$, and 
the statement follows from Theorem~\ref{prelim-thm2}.

(5)  Note that  the map $\cc^M\ni\alpha\mapsto\cL_{(\alpha)}$ is entire analytic. Since  $e(\alpha)$ is a simple eigenvalue of $\cL_{(\alpha)}$ for all
$\alpha\in\rr^M$, the regular perturbation theory implies that $e(\alpha)$ is  a real analytic function of $\alpha$. Property (2) and H\"older's 
inequality yield that $e(\alpha)$ is a convex function of $\alpha$.

(6) This part also follows from  regular perturbation theory. Fix $\alpha_0\in\rr^M$ and set
$$
\delta=\frac12\min\{e(\alpha_0)-\Re z\,|\,z\in\sp(\cL_{(\alpha_0)})\setminus\{e(\alpha_0)\}\}>0.
$$
If $\epsilon$ is small enough and $\alpha\in D_\epsilon=\{z\in\cc^M\,|\,|\alpha-\alpha_0|<\epsilon\}$, 
one has
$$
\langle\rho|\e^{t\cL_{(\alpha)}}(\one)\rangle
=\e^{te(\alpha)}\left(\langle\rho|M_\alpha\rangle\langle\mu_\alpha|\one\rangle
+O\left(\e^{t(-\delta+O(\epsilon))}\right)\right),
$$
where $e(\alpha)$, $M_\alpha$ and $\mu_\alpha$ are analytic functions of $\alpha$
such that $\langle\rho|M_\alpha\rangle\langle\mu_\alpha|\one\rangle
-\langle\rho|M_{\alpha_0}\rangle\langle\mu_{\alpha_0}|\one\rangle=O(\epsilon)$ and
$\langle\rho|M_{\alpha_0}\rangle\langle\mu_{\alpha_0}|\one\rangle>0$. It follows that there exists 
$\epsilon>0$ such that for $\alpha\in D_\epsilon$,
$$
\lim_{t\to\infty}\frac1t\log\langle\rho|\e^{t\cL_{(\alpha)}}(\one)\rangle=e(\alpha).
$$

(7) Let $\Theta$ be the time-reversal map. By Property~(5) of Theorem~\ref{thm-DB} one has
$$
\Theta\circ\cL_{(\alpha)}^\ast=\cL_{(1-\alpha)}\circ\Theta,
$$
for all $\alpha\in\rr^M$.
It follows that $\sp(\cL_{(\alpha)})=\sp(\cL_{(1-\alpha)})$ and hence $e(\alpha)=e(1-\alpha)$.

(8) If Hypothesis  (\KMSB) is satisfied, then $\rho_j=Z_j^{-1}\nu^{\beta_j}$ with $\nu=\e^{-H_\cS}$
and $Z_j=\tr(\nu^{\beta_j})$. Hence, 
$\cR_{\rho_j}^{\alpha_j}=Z_j^{-2\alpha_j}\cR_\nu^{\alpha_j\beta_j}$ and Part (4) of 
Theorem~\ref{thm-DB} yields
$$
\cL_{(\alpha)}=\sum_{j=1}^M\cR_\nu^{\alpha_j\beta_j/2}\circ\cL_j\circ\cR_\nu^{-\alpha_j\beta_j/2}.
$$
It follows that
\[\cL_{(\alpha+\lambda\beta^{-1})}=\cR_\nu^{\lambda/2}\circ\cL_{(\alpha)}\circ\cR_\nu^{-\lambda/2},\]
and so  $\sp(\cL_{(\alpha+\lambda\beta^{-1})})=\sp(\cL_{(\alpha)})$. In particular, 
$e(\alpha+\lambda\beta^{-1})=e(\alpha)$.

\subsection{Proof of Theorem~\ref{prop-Flux}}

\noindent(1)
At $\alpha=0$, the spectral projection of $\cL_{(\alpha)}=\cL$ associated to its dominant eigenvalue
$1$ is $|\one\rangle\langle\rho_+|$. Thus,  for $\alpha$ sufficiently close to $0\in\rr^M$, 
$e(\alpha)={E_1(\alpha)}/{E_0(\alpha)}$ where
$$
E_n(\alpha)=\oint_\Gamma z^n\langle\rho_+|(z-\cL_{(\alpha)})^{-1}(\one)\rangle\frac{\d z}{2\pi\i},
$$
and $\Gamma$ is a small circle centred at $1$ such that  no other point of $\sp(\cL)$ is 
on or inside $\Gamma$.
Since $(z-\cL)^{-1}(\one)=z^{-1}$, one has $E_1(0)=0$ and $E_0(0)=1$ and hence
$(\partial_{\alpha_j}e)(0)=(\partial_{\alpha_j}E_1)(0)$. An elementary calculation yields
$$
(\partial_{\alpha_j}E_n)(0)
=\oint_\Gamma z^n\langle\rho_+|(z-\cL)^{-1}\circ\cL_{;\alpha_j}\circ
(z-\cL)^{-1}(\one)\rangle\frac{\d z}{2\pi\i},
$$
where 
\beq
\cL_{;\alpha_j}(X)=\left.\partial_{\alpha_j}\cL_{(\alpha)}(X)\right|_{\alpha=0}
=\cL_j(XS_j)-\cL_j(X)S_j.
\label{cLfd}
\eeq
The identities 
\[\cL_{;\alpha_j}\circ(z-\cL)^{-1}(\one)=z^{-1}\cL_{;\alpha_j}(\one)=z^{-1}\cI_j,\]
\[\langle\rho_+|(z-\cL)^{-1}=z^{-1}\langle\rho_+|,\] yield
$$
(\partial_{\alpha_j}E_0)(0)=0,\qquad
(\partial_{\alpha_j}E_1)(0)=\rho_+(\cI_j),
$$
and the statement follows.

\noindent(2)
From the previous calculation, we easily infer
\begin{align}
(\partial_{\alpha_k}\partial_{\alpha_j}e)(0)&=(\partial_{\alpha_k}\partial_{\alpha_j}E_1)(0)\nonumber\\[4pt]
&=\oint_\Gamma\frac1z\langle\rho_+|\cL_{;\alpha_k}\circ(z-\cL)^{-1}(\cI_j)
+\cL_{;\alpha_j}\circ(z-\cL)^{-1}(\cI_k)
+\cL_{;\alpha_k\alpha_j}(\one)\rangle\frac{\d z}{2\pi\i},
\label{sunnybutfresh}
\end{align}
where 
\[\cL_{;\alpha_k\alpha_j}(\one)=\left.\partial_{\alpha_k}\partial_{\alpha_j}\cL_{(\alpha)}(\one)\right|_{\alpha=0}
=\delta_{kj}(\cL_j(S_j^2)-2\cL_j(S_j)S_j).\]
Theorem~\ref{thm-DB}~(2) implies
$$
\cL_j(\e^{tS_j}X\e^{-tS_j})=\e^{tS_j}\cL_j(X)\e^{-tS_j},
$$
and hence $\cL_j([S_j,X])=[S_j,\cL_j(X)]$. It follows that $[S_j,\cL_j(S_j)]=0$ and 
$\cL_{;\alpha_k\alpha_j}(\one)=\delta_{kj}D_j(S_j,S_j)$. Using  the fact that
$$
\oint_\Gamma\frac1z\langle\rho_+|\cL_{;\alpha_k}\circ(z-\cL)^{-1}(\rho_+(\cI_j))\rangle\frac{\d z}{2\pi\i}=
\oint_\Gamma\frac1{z^2}\rho_+(\cI_k)\rho_+(\cI_j)\frac{\d z}{2\pi\i}=0,
$$
we can replace  $\cI_{j/k}$ with $\cJ_{j/k}=\cI_{j/k}-\rho_+(\cI_{j/k})$ in Eq.~\eqref{sunnybutfresh}.
Since $\rho_+(\cJ_{j})=0$, the meromorphic function $(z-\cL)^{-1}(\cJ_{j})$ is regular at $z=0$ and 
one has
$$
\lim_{z\to0}(z-\cL)^{-1}(\cJ_{j})=\int_0^\infty\e^{t\cL}(\cJ_j)\,\d t,
$$
the integral on the r.h.s.\;being absolutely convergent. We therefore have
$$
(\partial_{\alpha_k}\partial_{\alpha_j}e)(0)
=\int_0^\infty
\rho_+(\cL_{;\alpha_k}(\e^{t\cL}(\cJ_j))+\cL_{;\alpha_j}(\e^{t\cL}(\cJ_k)))\,\d t
+\delta_{kj}\rho_+(D_j(S_j,S_j)).
$$
The relation 
\begin{align*}
\left.\frac{\partial^2 e(\alpha)}{\partial\alpha_j\partial\alpha_k}\right|_{\alpha=0}
&=-\int_0^\infty\rho_+\left(
\e^{t\cL}(\cJ_j)\cJ_k^++\e^{t\cL}(\cJ_k)\cJ_j^+\right)\d t\\
&+\int_0^\infty\rho_+\left(\cL_k(\e^{t\cL}(\cJ_j)S_k)+\cL_j(\e^{t\cL}(\cJ_k)S_j)\right)\d t
+\delta_{jk}\rho_+(D_j(S_j,S_j)),
\end{align*}
now follows from Eq.~\eqref{cLfd} and the identity
$$
\rho_+(\cL_{k}(\e^{t\cL}(\cJ_j))S_k)
=\langle\cL_{k}(\e^{t\cL}(\cJ_j))|S_k\rangle_{\rho_+}
=\langle\e^{t\cL}(\cJ_j)|\cL_{k}^{\rho_+}(S_k)\rangle_{\rho_+}
=\rho_+(\e^{t\cL}(\cJ_j)\cJ_k^+).
$$
Finally, an application of Vitali's convergence theorem (see Appendix B in \cite{JOPP}) gives 
\[
\left.\frac{\partial^2 e(\alpha)}{\partial\alpha_j\partial\alpha_k}\right|_{\alpha=0}
=\lim_{t\to\infty}t\langle(\varsigma_j-\langle\varsigma_j\rangle_{\rho,t})(\varsigma_k-\langle\varsigma_k\rangle_{\rho,t})\rangle_{\rho,t}.
\]


\begin{thebibliography}{999999}

\bibitem[Ag]{Ag} Agarwal, G.S.: 
\newblock Open quantum markovian systems and the microreversibility. 
\newblock Z. Physik {\bf 258}, 409-422 (1973). 

\bibitem[AGMT]{AGMT} Andrieux, D., Gaspard, P., Monnai, T., and Tasaki, S.:
\newblock The fluctuation theorem for currents in open quantum systems.
\newblock New J. Phys. {\bf 11}, 043014 (2009).

\bibitem[Al]{Al} Alicki, R.: 
\newblock On the detailed balance condition for non-Hamiltonian systems.
\newblock Rep. Math. Phys. {\bf 10}, 249--258 (1976).

\bibitem[AL]{AL} Alicki, R., and Lendi, K.: 
\newblock {\sl Quantum Dynamical Semigroups and Applications.}  
\newblock Springer, Berlin 2007.

\bibitem[AS]{AS} Aschbacher, W., and Spohn, H.:
\newblock A remark on the strict positivity of the entropy production.
\newblock Lett. Math. Phys. {\bf 75}, 17--23 (2006).

\bibitem[Br]{Br} Bryc, W.: 
\newblock A remark on the connection between the large deviation principle and the central limit theorem. 
\newblock Stat. Prob. Lett. {\bf 18}, 253--256 (1993). 

\bibitem[CE]{CE} Christensen, E., and Evans, D.E.:
\newblock Cohomology of operator algebras and quantum dynamical semigroups.
\newblock J. London Math. Soc. {\bf 20}, 358--368 (1979).

\bibitem[CM]{CM}Chetrite, R., and Mallick, K.:
\newblock Quantum fluctuation relations for the Lindblad master equation.
\newblock J. Stat. Phys. {\bf 148}, 480--501 (2012).

\bibitem[Da1]{Da0} Davies, E.B.:
\newblock Quantum Stochastic Processes II.
\newblock Commun. Math. Phys. {\bf 19}, 83--105 (1970).

\bibitem[Da2]{Da1}Davies, E.B.:
\newblock Markovian master equations.
\newblock Commun. Math. Phys. {\bf 39}, 91--110 (1974).

\bibitem[Da3]{Da2}Davies, E.B.:
\newblock Markovian master equations. II.
\newblock Math. Ann. {\bf 219}, 147--158 (1976).

\bibitem[Da4]{Da3}Davies, E.B.:
\newblock Markovian master equations. III.
\newblock Ann. Inst. H. Poincar\'e, section B, {\bf 11}, 265--273 (1975).

\bibitem[Da5]{Da4}Davies, E.B.:
\newblock {\sl Quantum Theory of Open Systems.} 
\newblock Academic Press, London (1976).

\bibitem[DdRM]{DdRM} Derezi\'nski, J., de Roeck, W., and Maes, C.:
\newblock Fluctuations of quantum currents and unravelings of master equations.
\newblock J. Stat. Phys. {\bf 131}, 341--356 (2008).

\bibitem[DF]{DF} Derezi\'nski, J., and Fr\"uboes, R.:
\newblock Fermi golden rule and open quantum systems. 
\newblock In {\sl Open Quantum Systems III. The Markovian Approach.}
\newblock S. Attal, A. Joye and C.-A. Pillet editors. 
\newblock Lecture Notes in Mathematics {\bf 1882}, Springer, Berlin, 2006.

\bibitem[DJ]{DJ} Derezi\'nski, J., and  Jak\v si\'c, V.:
\newblock  On the nature of Fermi Golden Rule for open quantum systems.
\newblock  J. Stat. Phys. {\bf 116},  411-423, (2004).

\bibitem[DJP]{DJP} Derezi\'nski, J., Jak\v si\'c, V., and Pillet, C.-A.: 
\newblock Perturbation theory of $W^\ast$-dynamics, Liouvilleans and KMS-states.
\newblock Rev. Math. Phys. {\bf 15}, 447--489 (2003).

\bibitem[dR1]{dR1} de Roeck, W.:
\newblock Quantum fluctuation theorem: Can we go from micro to meso?
\newblock Comptes Rendus Phys. {\bf 8}, 674--683 (2007).

\bibitem[dR2]{dR2} de Roeck, W.:
\newblock Large deviation generating function for currents  in the Pauli-Fierz model.
\newblock Rev. Math. Phys. {\bf 21}, 549--585 (2009).

\bibitem[dRK1]{dRK1} de Roeck, W., and Kupianien, A.: 
\newblock 'Return to Equilibrium' for weakly coupled quantum systems: a simple polymer expansion.
\newblock Commun. Math. Phys. {\bf 305}, 797-826 (2011). 

\bibitem[dRK2]{dRK2} de Roeck, W., and Kupianien, A.: 
\newblock Approach to ground state and time-independent photon bound for massless spin-boson models.
\newblock Ann. H. Poincar\'e {\bf 14}, 253--311 (2013).

\bibitem[dRM]{dRM} de Roeck, W., and Maes, C.:
\newblock Steady state fluctuations of the dissipated heat for a quantum stochastic model.
\newblock Rev. Math. Phys. {\bf 18}, 619--653 (2006).

\bibitem[DS]{DS} Davies, E.B., and Spohn, H.:
\newblock Open quantum systems with time-dependent Hamiltonians and their linear response.
\newblock J. Stat. Phys. {\bf 19},  511--523 (1978).

\bibitem[D\"uSp]{D\"uSp} D\"umcke, R.,  and Spohn, H.:
\newblock The proper form of the generator in the weak coupling limit.
\newblock Z. Physik B {\bf 34}, 419--422 (1979).

\bibitem[DZ]{DZ} Dembo, A., and Zeitouni, O.: 
\newblock {\em Large Deviations Techniques and Applications.} 
\newblock Second edition. Applications of Mathematics, {\bf 38}. 
Springer, New York (1998). 

\bibitem[ECM]{ECM} Evans, D.J., Cohen, E.G.D., and Morriss, G.P.:
\newblock Probability of second law violation in shearing steady flows. 
\newblock Phys. Rev. Lett. {\bf 71}, 2401--2404 (1993).

\bibitem[EHM]{EHM} Esposito, M., Harbola, U., and Mukamel, S.:
\newblock Nonequilibrium fluctuations, fluctuation theorems, and counting
statistics in quantum systems.
\newblock Rev. Mod. Phys. {\bf 81}, 1665--1702 (2009).

\bibitem[EHK]{EHK} Evans, D.E., and H\o egh-Krohn, R.:
\newblock Spectral properties of positive maps on $C^\ast$-algebras.
\newblock J. London. Math. Soc. {\bf 17}, 345--355 (1978).

\bibitem[El]{El} Ellis, R.S.:
\newblock{\em Entropy, Large Deviations, and Statistical Mechanics.}
\newblock Springer, Berlin (1985). Reprinted in the series Classics of Mathematics (2006).

\bibitem[ES]{ES} Evans, D.J., and Searles, D.J.: 
\newblock Equilibrium microstates which generate second law violating steady states. 
\newblock Phys Rev. E {\bf 50}, 1645--1648 (1994).

\bibitem[FU]{FU} Fagnola, F., and Umanit\`a, V.:
\newblock Detailed balance, time reversal, and generators of quantum Markov semigroups.
\newblock Mathematical Notes {\bf 84}, 108--115 (2008).

\bibitem[Ga]{Ga} Gallavotti, G.: 
\newblock Extension of Onsager's reciprocity to large fields and the chaotic hypothesis. 
\newblock Phys. Rev. Lett. {\bf 77}, 4334--4337 (1996).

\bibitem[GC1]{GC1} Gallavotti, G., and Cohen, E.G.D.: 
\newblock Dynamical ensembles in nonequilibrium statistical mechanics.
\newblock Phys. Rev. Lett. {\bf 74}, 2694--2697 (1995).

\bibitem[GC2]{GC2} Gallavotti, G., and Cohen, E.G.D.: 
\newblock Dynamical ensembles in stationary states. 
\newblock J. Stat. Phys. {\bf 80}, 931--970 (1995).

\bibitem[GKS]{GKS} Gorini, V., Kossakowski, A., and Sudarshan, E.C.G.:
\newblock Completely positive dynamical semigroups of N-level systems. 
\newblock J. Math. Phys. {\bf 17}, 821--825 (1976).

\bibitem[JOPP]{JOPP} Jak\v si\'c, V., Ogata, Y., Pautrat, Y., and Pillet, C.-A.:
\newblock Entropic fluctuations in quantum statistical mechanics -- an introduction.
\newblock In {\sl Quantum Theory from Small to Large Scales.} J.~Fr\"ohlich,  M.~Salmhofer,
W.~de~Roeck, V.~Mastropietro and L.F.~Cugliandolo editors. Oxford University Press, Oxford, 2012.

\bibitem[JOPS]{JOPS} Jak\v si\'c, V., Ogata, Y., Pillet, C.-A., and Seiringer, R.:
\newblock Quantum hypothesis testing and non-equilibrium statistical mechanics.
\newblock Rev. Math. Phys. {\bf 24}, 1230002 (2012).

\bibitem[JP1]{JP1}  Jak\v si\'c, V., and Pillet C.-A.:
\newblock  Non-equilibrium steady states of finite quantum 
systems coupled to thermal reservoirs.
\newblock  Commun. Math. Phys. {\bf 226},  131-162 (2002).

\bibitem[JP2]{JP} Jak\v si\'c, V., and Pillet C.-A.:  
\newblock Entropic functionals in quantum statistical mechanics.
\newblock To appear in {\sl Proceedings of the XVIIth International Congress of Mathematical Physics, Aalborg, Denmark, 2012.}

\bibitem[JPR]{JPR} Jak\v si\'c, V., Pillet C.-A, Rey-Bellet, L.: 
\newblock Entropic fluctuations in statistical mechanics I. Classical dynamical systems. 
\newblock Nonlinearity {\bf 24}, 699 (2011).

\bibitem[JPPW]{JPPW} Jak\v si\'c, V., Panati, A.,  Pillet, C.-A.,  and Westrich, M.:
\newblock Non-equilibrium statistical mechanics of Pauli-Fierz systems.
\newblock In preparation.

\bibitem[KFGV]{KFGV} Kossakowski, A., Frigerio, A., Gorini, V., and Verri, M.:
\newblock Quantum detailed balance and KMS condition.
\newblock Commun. Math. Phys. {\bf 57}, 91--110 (1977).

\bibitem[Ku1]{Ku1} Kurchan, J.: 
\newblock Fluctuation theorem for stochastic dynamics.  
\newblock J. Phys. A  {\bf 31}, 3719--3729 (1998).

\bibitem[Ku2]{Ku2} Kurchan, J.:
\newblock A Quantum Fluctuation Theorem.
\newblock Preprint 2000 (arXiv: cond-mat/0007360).

\bibitem[Lb]{Lb} Lieb, E.H.: 
\newblock Convex trace functions and the Wigner-Yanase-Dyson conjecture.
\newblock Adv. in Math. {\bf 11}, 267--288 (1973).

\bibitem[Li1]{Li1} Lindblad, G.: 
\newblock On the generators of quantum dynamical semigroups.
\newblock Commun. Math. Phys. {\bf 48}, 119--130 (1976).

\bibitem[Li2]{Li2} Lindblad, G.: 
\newblock Dissipative operators and cohomology of operator algebras.
\newblock Lett. Math. Phys. {\bf 1},  219--224 (1976).

\bibitem[LS1]{LS1} Lebowitz, J.L., and Spohn, H.: 
\newblock Irreversible thermodynamics for quantum systems weakly 
coupled to thermal reservoirs.
\newblock Adv. Chem. Phys. {\bf 38}, 109--142 (1978).

\bibitem[LS2]{LS2} Lebowitz, J.L., and Spohn, H.: 
\newblock A Gallavotti-Cohen-type symmetry in the large deviation functional 
for  stochastic dynamics.  
\newblock J. Stat. Phys. {\bf 95}, 333--365  (1999).

\bibitem[M]{M} Maes, C.:
\newblock The fluctuation theorem as a Gibbs property. 
\newblock J. Stat. Phys. {\bf 95}, 367--392 (1999).

\bibitem[Ma]{Ma} Majewski, W.A.:
\newblock The detailed balance condition in quantum statistical mechanics.
\newblock J. Math. Phys. {\bf 25}, 614--616 (1984).

\bibitem[MRV]{MRV} Maes, C.,  Redig, F., and Verschuere, M.: 
\newblock From global to local fluctuation theorems. 
\newblock Mosc. Math. J.  {\bf 1}, 421--438 (2001).

\bibitem[MT]{MT} Matsui, T., and Tasaki, S.: 
\newblock Fluctuation theorem, nonequilibrium steady states and MacLennan-Zubarev ensembles
of a class of large quantum systems.
\newblock In {\sl Fundamental Aspects of Quantum Physics,} L.~Accardi and  S.~Tasaki editors.
\newblock World Scientific, Singapore, 2003.

\bibitem[OP]{OP} Ohya, M., and Petz, D.: 
\newblock {\em  Quantum Entropy and its Use.} Second edition.
\newblock Springer,  Berlin, 2004.

\bibitem[Re]{Re} Rebolledo, R.:
\newblock Complete positivity and the Markov structure of open quantum systems.
\newblock In {\sl Open Quantum Systems II. The Markovian Approach.}
\newblock S. Attal, A. Joye and C.-A. Pillet editors. 
\newblock Lecture Notes in Mathematics {\bf 1882}, Springer, Berlin, 2006.

\bibitem[RM]{RM} Rondoni, L. and Mej\'\myi a-Monasterio, C.:
\newblock Fluctuations in non-equilibrium statistical mechanics: models, mathematical 
theory, physical mechanisms. 
\newblock Nonlinearity {\bf 20}, 1--37 (2007).

\bibitem[Schr]{Sc} Schrader, R.: 
\newblock Perron-Frobenius theory for positive maps on trace ideals.
\newblock In {\sl Mathematical physics in mathematics and physics (Siena, 2000)}, 361--378, 
Fields Inst. Commun., 30, AMS  Providence, RI, 2001.

\bibitem[Sp1]{Sp1} Spohn, H.:
\newblock Approach to equilibrium for completely positive dynamical semigroups of N-level systems.
\newblock Rep. Math. Phys. {\bf 10}, 189--194 (1976).

\bibitem[Sp2]{Sp2} Spohn, H.:
\newblock An algebraic condition for the approach to equilibrium of an open N-level system.
\newblock Lett. Math. Phys. {\bf 2}, 33--38 (1977).

\bibitem[Sp3]{Sp3} Spohn, H.:
\newblock Entropy production for quantum dynamical semigroups.
\newblock J. Math. Phys. {\bf 19}, 1227--1230 (1978).

\bibitem[St]{St} Stinespring, W.F.:
\newblock Positive functions on $C^\ast$-algebras. 
\newblock Proceedings of the American Mathematical Society {\bf 6}, 211--216 (1955).

\bibitem[Uh]{Uh} Uhlmann, A.:
\newblock Relative entropy and the Wigner-Yanase-Dyson-Lieb concavity in an interpolation theory.
\newblock Comm. Math. Phys. {\bf 54}, 21--32 (1977).
\end{thebibliography}
\end{document}